\documentclass[structabstract]{aa}  
\usepackage{graphicx}
\usepackage{txfonts}
\usepackage[round]{natbib}
\def \kh{\langle\kappa\rangle_{\mathrm{H}}}
\def \kcrit{\kappa_{\mathrm{crit}}}
\def \kap{\kappa_{\mathrm{acc}}}
\def \qq{Q_{\mathrm{acc}}}
\def \ag{a_{\mathrm{gr}}}

\begin{document}
   \title{Exploring wind-driving dust species in cool luminous giants}

   \subtitle{I. Basic criteria and dynamical models of M-type AGB stars}

   \author{S. Bladh
          \inst{1}
          \and
          S. H\"ofner\inst{1}
          }
   \institute{Department of Physics and Astronomy, Division of Astronomy and Space Physics, Uppsala University,
              Box 516, SE-75120, Uppsala, Sweden\\
              \email{sara.bladh@fysast.uu.se}
             }

   \date{Received February 29, 2012; accepted September 5, 2012}

 
  \abstract
     {The heavy mass loss observed in evolved AGB stars is usually attributed to a two-stage process: atmospheric levitation by pulsation-induced shock waves followed by radiative acceleration of dust grains, which transfer momentum to the surrounding gas through collisions. In order for an outflow to occur the two stages of the mass-loss scheme have to connect,  i.e. the radiative acceleration can only be initiated if the levitated gas reaches a distance from the stellar photosphere where dust particles can condense. This levitation distance is limited by the kinetic energy transferred to the gas by the shock waves, which imposes strict constraints on potential wind-driving dust species.}
   {This work is part of an ongoing effort aiming at identifying the actual wind-drivers among the dust species observed in circumstellar envelopes. In particular, we focus on the interplay between a strong stellar radiation field and the dust formation process.}
   {To identify critical properties of potential wind-driving dust species we use detailed radiation-hydrodynamical models which include a parameterized dust description, complemented by simple analytical estimates to help with the physical interpretation of the numerical results. The adopted dust description is constructed to mimic different chemical and optical dust properties in order to systematically study the effects of a realistic radiation field on the second stage of the mass loss mechanism.}
   {We see distinct trends in which combinations of optical and chemical dust properties are needed to trigger an outflow. Dust species with a low condensation temperature and a NIR absorption coefficient that decreases strongly with wavelength will not condense close enough to the stellar surface to be considered as potential wind-drivers.}
   {Our models confirm that metallic iron and Fe-bearing silicates are not viable as wind-drivers due to their near-infrared optical properties and resulting large condensation distances. TiO$_2$ is also excluded as a wind-driver due to the low abundance of Ti. Other species, such a SiO$_2$ and Al$_2$O$_3$, are less clear-cut cases due to uncertainties in the optical and chemical data and further work is needed. A strong candidate is Mg$_2$SiO$_4$ with grain sizes of 0.1-1 $\mu$m, where scattering contributes significantly to the radiative acceleration, as suggested by earlier theoretical work and supported by recent observations.}

   \keywords{  Stars: AGB and post-AGB Ð Stars: mass-losss Ð Stars: atmospheres Ð Stars: winds, outflows, circumstellar matter, dust Ð Stars: Hydrodynamics, Radiative transfer
               }

   \maketitle
\section{Introduction}
\label{s_intro}
The prevailing scenario for stellar winds observed in AGB stars is a two-stage process, involving atmospheric levitation by shock waves and radiative acceleration of dust grains. In the first step pulsation-induced shock waves catapult material outwards in almost ballistic trajectories. The pulsations themselves are not sufficient to cause an outflow that is comparable with observed mass loss rates or velocities \citep[e.g.][]{woo79,b88} but the shocks typically manage to lift the gas to a few stellar radii. As a consequence of the propagating shock waves in the atmosphere, density-enhanced layers of gas form around the star; layers where the temperature is low enough for dust formation to occur {\citep[e.g][]{flei92,jeong03}. The dust is of crucial importance in the second step of the mass loss scheme; if dust grains with adequate extinction cross-sections condense, the momentum gained from photons which are absorbed or scattered by these particles may be sufficient to overcome the gravitational pull of the star and accelerate the grains outwards. The collisions between dust particles and the surrounding gas cause a transfer of momentum that triggers a general outflow. 
   
Observations tell us that there exists a number of different dust species in the circumstellar environment  \citep[see, e.g.,][for an overview]{agbgrain}, however, most of these are not abundant enough or have insufficient cross-sections to generate the radiative acceleration needed to drive a wind. Furthermore, in order for a dust species to trigger a wind it has to form close to the stellar surface, within reach of the shock waves. Due to an environment dominated by a strong radiation field the closest distance from the star at which a given dust species can exist will depend on both the optical and chemical properties of the material.

The chemical composition of dust grains in the circumstellar environments of AGB stars is first and foremost determined by the elemental abundances in the atmosphere. As stars evolve during the AGB phase their atmospheric chemistry may transform from oxygen-dominated (M-type, C/O $<1$) to carbon-rich (C-type, C/O $>1$) due to ongoing nuclear burning and dredging-up of processed material  \citep[see, e.g., the review by][]{herw05}. Since most of the carbon and oxygen in the extended atmosphere will be tied up in the tightly bound CO-molecule, whatever element is more abundant will dominate the dust chemistry. As a consequence, carbon-based condensates are expected to be the main wind-drivers in C-type AGB stars. Self-consistent radiation-hydrodynamical models (RHD models) reproduce well observed wind properties and spectra  of C-type AGB stars  \citep[e.g.][]{berwin98,wint00,and03,gloidl04,now10,now11,sac11}, which demonstrates convincingly that the outflows are driven by radiative acceleration of amorphous carbon grains. 

For M-type AGB stars, on the other hand, where the dust chemistry is much more complex, it is still a matter of debate which grain species are responsible for driving the outflows \citep[see, e.g., the discussion in][]{hof09}. Observations of mid-infrared spectral features indicate that magnesium-iron silicates (olivine [Mg,Fe]$_2$SiO$_4$ and pyroxene [Mg,Fe]SiO$_3$) are abundant in the circumstellar environment \citep[e.g.][]{agbgrain}. However, the Mg/Fe ratio of the grain material is poorly constrained by such observations. The grain composition is important because it strongly influences the absorption properties in the near-infrared, where the star emits most of its radiation, and consequently affects the resulting radiative acceleration. Fe-free silicates are extremely transparent in the near-infrared but the absorption cross-section increases dramatically with Fe content.

Chemical modeling based on phase-equilibrium between gas and dust favors silicate grains at the Mg-rich end of the spectrum, but the assumption of chemical equilibrium is not justified in the highly dynamical atmospheres of AGB stars. Kinetic models of grain growth, on the other hand, point towards roughly equal amounts of Mg and Fe, reflecting the relative abundances of these elements \citep[see, e.g.,][for a detailed discussion on modeling of dust formation]{gailminerology}. The complex interplay of the stellar radiation field with the grain growth process, however, has a decisive influence on the grain composition since the grain temperature at a given distance from the stellar surface is strongly affected by the Fe content.

Using frequency-dependent wind models with a detailed treatment of dust formation, \citet{woi06fe} demonstrated that silicate grains have to be virtually Fe-free at distances corresponding to the wind acceleration zone. But the lack of Fe leads to low absorption cross-sections in the near-infrared region, resulting in insufficient radiative acceleration to produce outflows. \citet{hof08bg} suggested scattering as a possible solution: if conditions allow Fe-free silicate grains to grow into a size range of about $0.1-1\mu$m, the scattering cross-section becomes dominant over absorption by several orders of magnitude, opening up the possibility of stellar winds driven by photon scattering. The models by \citet{hof08bg}, which include a detailed description for the growth of Mg$_2$SiO$_4$ grains, show mass loss rates and wind velocities typical of AGB stars, as well as realistic photometric properties \citep[cf.][]{bladh11}. Strong observational support for the scenario of winds driven by photon scattering comes from recent work by \cite{norr12}, who detected dust particles of sizes $\sim0.3~\mu$m in the close circumstellar environment of three M-type AGB stars, using multi-wavelength aperture-masking polarimetric interferometry.\footnote{The fraction of light scattered by dust is modeled assuming a geometrically thin shell of Fe-free silicates (forsterite) and a uniform grain size. The wavelength-dependence of the scattering cross-section allows them to derive the grain size from their multi-wavelengths measurements.} This answers the question whether grains can grow sufficiently large in the extended atmosphere.

Fe-free silicates are clearly a possible candidate for driving winds in M-type AGB stars but the question remains if other dust species can trigger winds or contribute to accelerating outflows. At present it is not clear whether winds are driven by a single type of grain material, a combination of several dust species or composite grains consisting of mixed materials (i.e. dirty grains or core mantel grains). In addition to silicates, observations show features attributed to other dust species such as spinel, corundum and simple oxides.

Previous theoretical work on dust chemistry in AGB stars has focused on detailed descriptions of grain growth, investigating the time-dependent formation of homogenous and mixed grains, as well as networks of dust species competing for the same raw materials \citep[e.g.][]{gail88,gaug90,gail99,fega01,jeong03,hewo06,woi06fe}. Most existing studies on dust-driven winds of M-type AGB stars \citep[e.g][]{jeong03,fega06}, however, are based on an crude treatment of the radiation field, resulting in unrealistic gas and grain temperatures and potentially misleading conclusions about the dust composition, as illustrated by the example of the Mg/Fe-ratio for silicate grains discussed above.

The work presented here is part of an ongoing effort to investigate how a strong radiation field affects dust condensation. To facilitate such an investigation we consciously scale back on the complexity of the dust properties and the grain growth process by using a parameterized formula for the dust opacity in our frequency-dependent RHD models. This formula captures the essential optical and chemical properties of the grain material when considering the interplay between the dust component and the radiation field. By constructing a grid of models where we systematically vary these properties, instead of conducting detailed studies for a few selected dust species, we get an overall picture of how the interaction between the radiation field and the dust component affects the dynamical properties of the model atmospheres. This knowledge makes it possible to exclude dust species which are not viable as wind-drivers and to identify possible candidates for which a more detailed modeling is required.

In the current paper we focus on the dynamics of the atmospheres and winds. In a forthcoming paper we will give additional constraints based on comparisons of synthetic and observed spectra and photometric data. The paper is organized in the following way: In Sect. \ref{s_toy} we introduce simple analytical estimates to pinpoint the criteria that have to be fulfilled for a dust species to be considered as a potential wind-driver. Sect. \ref{s_model} includes a description of the RHD wind model and the parameterized dust opacity. In Sect. \ref{s_mpid} we present the setup of the computational grid and how the input parameters were chosen. A presentation of the numerical results is given in Sect. \ref{s_res} and we comment on specific dust species in Sect. \ref{s_spg}. In Sect. \ref{s_concl} we provide a summary of our conclusions. 

\section{Dust driven winds: a qualitative picture}
\label{s_toy}
Given the importance of dust in the mass loss mechanism of AGB stars it is essential to understand how the close stellar environment affects dust formation processes, and which grain properties are crucial to produce a mass outflow from the star. Some simple analytical constructs can help us gain qualitative insights into what properties of a grain material determine if a dust species is a possible wind-driver or not, paving the way for a physical interpretation of detailed RHD models.

\subsection{Radiative acceleration of the wind}
\label{s_radacc}
Consider the dynamics of a fluid element, starting right after it has been accelerated outwards by a shock wave: the main forces acting on the fluid are the gravitational force and the radiative acceleration on the dust particles within the fluid element. The force due to the thermal pressure gradient in the atmosphere is small by comparison and we neglect it in this simple construct. It follows that the acceleration of the fluid element in the radial direction is mainly controlled by the relative value of the gravitational and radiative acceleration, which leads to the following equation of motion,
\begin{equation}
\label{e_toy}
\frac{du}{dt}=-a_{\mathrm{grav}}+a_{\mathrm{rad}}=-a_{\mathrm{grav}}\left(1-\Gamma\right)=-g_*\left(\frac{R_*}{r}\right)^2\left(1-\Gamma\right),
\end{equation}
assuming a direct coupling between the gas and the dust components. Here $u=dr/dt$ is the radial velocity of the fluid element at distance $r$ from the stellar center, $g_* = GM_*/R_*^2$ denotes the gravitational acceleration at the stellar radius $R_*$ and $\Gamma$ is a dimensionless quantity, dependent on $r$, that measures the ratio between radiative and gravitational acceleration:
\begin{equation}
\label{e_gamma}
\Gamma(r) = \frac{a_{\mathrm{rad}}}{a_{\mathrm{grav}}}=\frac{\kh L_*}{4\pi cGM_*}.
\end{equation}
Included in the  expression for $\Gamma$ is the gravitational constant $G$, the speed of light $c$, the stellar mass and luminosity, $M_*$ and $L_*$ respectively, and the total flux-averaged dust opacity $\kh$,
\begin{equation}
\label{e_kh}
\kh=\frac{\int_0^{\infty}\kap(\lambda)F_{\lambda}d\lambda}{\int_0^{\infty}F_{\lambda}d\lambda},
\end{equation}
where $F_{\lambda}$ is the monochromatic flux at wavelength $\lambda$ and $\kap$ is the monochromatic grain opacity (see below). We define a critical value of the flux-averaged opacity when the gravitational and radiative acceleration are in balance, i.e. when $\Gamma=1$.
\begin{equation}
\label{e_kcrit}
\kcrit= \frac{4\pi cGM_*}{L_*}
\end{equation}
Clearly the fluid element will accelerate outwards if $\Gamma>1$ (i.e. if $\kh > \kcrit$) and a closer examination of this quantity can help us pinpoint the grain properties necessary for driving a wind.  The expression for $\Gamma$ can be factorized into two parts; one part that solely depends on stellar parameters and fundamental physical constants, $L_*/4\pi cGM_*$, and one part, $\kh$, that is grain material dependent. Assuming for simplicity that all grains in the fluid element have equal radii $\ag$, their collective opacity per mass of stellar matter can be expressed as
\begin{equation}
\label{e_kap1}
\kap(\lambda,\ag)=\frac{\pi}{\rho} \ag^2\qq(\lambda,\ag)n_{\mathrm{gr}},
\end{equation}
where $n_{\mathrm{gr}}$ is the number density of grains and the efficiency $\qq$ is defined as the ratio between the radiative and the geometrical cross-section of an individual grain, which can be computed from optical data using Mie theory \citep[e.g.][]{bohuff83}. Introducing the atomic weight of the monomer $A_{\mathrm{mon}}$ (the basic building block of the grain material), the bulk density of the grain material $\rho_{\mathrm{gr}}$, abundances of the limiting element $\varepsilon_{\mathrm{lim}}$ and of helium $\varepsilon_{\mathrm{He}}$, as well as the degree of condensation of the limiting element $f_{\mathrm{c}}$ (the fraction of the limiting element condensed into grains), the expression for the dust opacity can be reformulated as (see Appendix \ref{a_radacc} for details)
\begin{equation}
\label{e_kap2}
\kap(\lambda,\ag)=\frac{3}{4}\frac{A_{\mathrm{mon}}}{\rho_{\mathrm{gr}}}\frac{\qq(\lambda,\ag)}{\ag}\frac{\varepsilon_\mathrm{lim}}{s(1+4\varepsilon_{\mathrm{He}})} f_\mathrm{c}.
\end{equation}
In this context we define the limiting element as the first element that will be completely consumed due to the relative element abundances in the atmosphere, adjusted for the number of atoms contributing to the monomer, $s$ (stoichiometric coefficient). For example, in the case of Mg$_2$SiO$_4$, the least abundant element would be Si (for a solar composition) but since 2 Mg atoms are used for building one monomer ($s=2$), Mg becomes the limiting element.\footnote{Note that this is a simple way of estimating the maximum amount of condensable material for a specific dust species, and consequently, an upper limit for the opacity. This, however, does not necessarily mean that the addition of this element corresponds to the slowest rate (bottle-neck) in building up a monomer.} The factor $A_{\mathrm{mon}}/\rho_{\mathrm{gr}}$ is a material-dependent constant and the optical properties of the dust particles are captured in the efficiency $\qq$, which includes contributions from both absorption and scattering

\begin{equation}
\label{e_qtot}
\qq= Q_{\mathrm{abs}} + (1-g_{\mathrm{sca}})Q_{\mathrm{sca}}
\end{equation}
($g_{\mathrm{sca}}$ is the asymmetry factor describing deviations from isotropic scattering). In the small particle limit, where the grain radius is much smaller than the relevant wavelengths, the absorption and scattering efficiencies behave like $Q_{\mathrm{abs}}\propto \ag$ and $Q_{\mathrm{sca}}\propto \ag^4$, according to Mie theory \citep[e.g.][]{bohuff83}. In this limit absorption dominates over scattering, implying that $\qq\approx Q_{\mathrm{abs}}$ and consequently that the efficiency per grain radius, $\qq/\ag\approx Q_{\mathrm{abs}}/\ag$, becomes independent of grain size. We can therefore express the grain opacity during the early stages of grain formation, or during the whole process if particles remain small, as a purely wavelength-dependent function,
\begin{equation}
\label{e_kapspl}
\kap(\lambda)=\frac{3A_{\mathrm{mon}}}{4\rho_{\mathrm{gr}}}\cdot \qq^{\prime}(\lambda)\cdot\frac{\varepsilon_\mathrm{lim}}{s(1+4\varepsilon_{\mathrm{He}})}\cdot f_\mathrm{c} \quad (2\pi \ag\ll\lambda)
\end{equation}
where $\qq^{\prime}=Q_{\mathrm{abs}}/\ag$. Looking at each of the factors in this expression individually reveals what conditions need to be satisfied for a dust species to be a potential wind-driver. First, the efficiency per grain radius $\qq^{\prime}$ has to be sufficiently large in the wavelength region of the stellar flux maximum, given that $\kh$ is calculated by taking the flux-mean of the grain opacity $\kap$. Furthermore, the abundance $\varepsilon_\mathrm{lim}$ is a limiting factor in the dust formation process. The last factor in the expression for the grain opacity, the degree of condensation $f_\mathrm{c}$, is a measure of how much of the limiting element has actually condensed into solid material. The degree of condensation will remain zero until the grains start to condense and grow. For this to happen, the gas has to be sufficiently cool and the grains have to be thermally stable. The grains are heated by the stellar radiation and consequently their temperature depends on both the optical properties of the grain material and the stellar flux distribution, as well as the distance from the star. The grain formation process is therefore strongly affected by the radiation field in the close stellar environment and the effect on grain temperature has to be taken into account when considering possible wind-driving dust species.

\begin{table*}
\caption{Properties of a few selected dust species. Columns 2-7 list grain material properties, i.e. the atomic weight of the monomer $A_{\mathrm{mon}}$, the bulk density $\rho_{\mathrm{gr}}$, the limiting element and the adjusted abundance of said element $\varepsilon_{\mathrm{lim}}/s$, the power law coefficient $p$ and the condensation temperature $T_{\mathrm{c}}$. Column 8-10 include properties that depend on the chosen stellar parameters (see Sect. \ref{s_mpid}), i.e. the condensation distance $R_{\mathrm{c}}$, the flux-averaged dust opacity for full condensation of the limiting element $\kh^{\mathrm{max}}$ and $\Gamma$. }             
\label{t_gprop}     
\centering                     
\begin{tabular}{l c c c c r c c c c l}        
\hline\hline                 
Material & $A_{\mathrm{mon}}$ & $\rho_\mathrm{{gr}}$ & lim. & $\varepsilon_{\mathrm{lim}}/s$ & $p$ & $T_{\mathrm{c}}$  &  $R_{\mathrm{c}}$ &  $\kh^{\mathrm{max}}$  & $\Gamma$ & References \& comments\\    
              & $(u) $ & (g/cm$^3$) & element & & & (K) & ($R_*$) &  (cm$^2$/g)  & & \\   
\hline       
   Fe & 56 & 7.87   & Fe & $3.24\cdot 10^{-5}$ & 2.4 & 1050 & 11.5 & 1.5 & 0.6 &1, 4. \\  
   Al$_2$O$_3$ & 102 & 3.97 & Al & $1.48\cdot 10^{-6}$ & 1.7 & 1400 & 3.6 & $4\cdot  10^{-2}$ & $2\cdot 10^{-2}$ & 1, 6, 10. \\
   TiO$_2$ & 80 & 4.26 & Ti & $9.77\cdot 10^{-8}$ & 0.1 & 1100  & 3.4 & $6\cdot  10^{-5}$ & $2\cdot 10^{-5}$ & 2, 3. \\
   SiO$_2$ & 60 & 2.20  & Si & $3.55\cdot 10^{-5}$ & 0.2 & 1050 & 3.9 & $6\cdot  10^{-2}$ & $2\cdot 10^{-2}$ & 1, 9. Interpolated optical data.\\ 
   MgSiO$_3$ & 100 & 2.71    & Si & $3.55\cdot 10^{-5}$ & $-0.5$ & 1100 & 2.6 & $5\cdot 10^{-2}$ & $2\cdot 10^{-2}$  & 1, 7. \\  
   Mg$_2$SiO$_4$ & 140 & 3.27    & Mg & $1.90\cdot  10^{-5}$ & $-0.9$ & 1100 & 2.1  & $3\cdot 10^{-2}$ & $1\cdot 10^{-2}$  & 1, 7. SPL in col. 6, 9-10.\\
   MgFeSiO$_4$ & 172 & 3.75 & Fe & $3.24\cdot 10^{-5}$ & 2.3 &  1100 & 9.5  & 2.9 & 1.1 & 1, 8. \\ 
\hline   
   amC & 12 & $ 1.85 $ & C & $1.85\times 10^{-4}$ &1.2 & 1700 & 1.8 & 6.9 & 2.6 & 2, 5. C-rich atmospheres.\\    
\hline                             
\end{tabular}
\tablebib{(1)~\citet{gailminerology};
(2) \citet{latt78}; (3) \citet{zeid11}; (4) \citet{ord88fe};
(5) \citet{roul91amc}; (6) \citet{koi95cor}; (7) \citet{jag03}:
(8) \citet{dor95}: (9) \citet{pal85}:
(10) \citet{beg97}.
}
\tablefoot{For the abundances of the limiting elements we adopted the values from \cite{andgrev89}, except for C, N and O where we
took the data from \cite{saugrev94}. For amC the abundance was calculated using the formula $\varepsilon_{\mathrm{C}}=(\mathrm{C/O}-1)\varepsilon_{\mathrm{O}}$, with a C/O-ratio of 1.25. The first reference in column 11 corresponds to condensation temperature (from chemical equilibrium calculations) and the second to bulk density and optical data in the NIR. The values given for SiO$_2$ in the NIR are derived from an interpolation, using data from \citet{pal85}, and probably overestimating the cross-sections. SPL is short for small particle limit.}
\end{table*}

\subsection{Condensation and levitation distances}

\label{s_rcond}
\begin{figure}
\centering
\includegraphics[width=8.5cm]{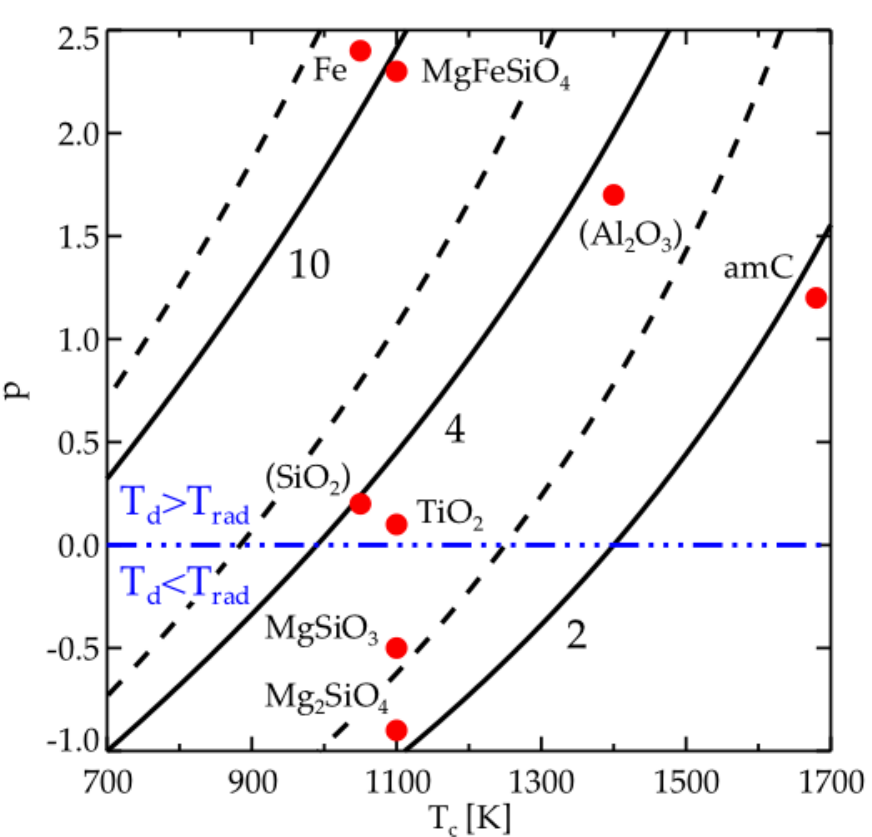}
   \caption{Curves of constant condensation distance ($R_{\mathrm{c}}/R_*=2,4,10$) as a function of the power law coefficient $p$ and the condensation temperature $T_{\mathrm{c}}$, using Eq. (\ref{e_rcond}), with $T_*=2800\,$K (solid lines) and $T_*=2500\,$K (dashed lines). The filled red circles mark where the selected set of dust species are situated in the $p/T_{\mathrm{c}}$-plane. The dust species in parenthesis have uncertain or interpolated optical data in the near-infrared. For $T_{\mathrm{c}}$ and $p$ values for the individual dust species, see Tab. \ref{t_gprop}}
   \label{f_rcond}
\end{figure}
In order to estimate if a specific grain material can start to condense in the density-enhanced layers of the atmosphere, we introduce the concepts of levitation distance and condensation distance, and some easy-to-use approximations. In the following, the levitation distance $R_{\ell}$ is defined as the distance to which pulsation-induced shock waves levitate the gas in the atmosphere, without radiative acceleration on dust. The condensation distance $R_{\mathrm{c}}$ is defined as the closest distance to the star were grains of a specific type can exist, i.e. are thermally stable. For a grain material to be considered a possible wind-driver, the two stages of the mass-loss scheme have to connect. The second stage, i.e. the radiative acceleration, can only be initiated if levitation by shock waves lifts gas beyond the condensation distance.

A simple argument using the complete conversion of pulsation-induced kinetic energy into potential energy, ignoring heat loss and pressure effects, can give us an estimate of how high shock waves can lift gas. If we assume the gas has an initial velocity $u_{\mathrm{0}}$ at distance $R_0$ from the center of the star we can derive the following expression for the levitation distance
\begin{equation}
\label{e_rmax}
\frac{R_{\ell}}{R_*} = \frac{R_0}{R_*}\left[1- \frac{R_0}{R_*}\left(\frac{u_{\mathrm{0}}}{u_{\mathrm{esc}}}\right)^2\right]^{-1},
\end{equation}
where the escape velocity at the stellar surface is given by  $u_{\mathrm{esc}}=(2M_*G/R_*)^{1/2}$ (see Appendix \ref{a_levd}). Given stellar parameters typical of an AGB star, i.e. $M_* = 1\,M_{\odot}$, $L_* = 5000\,L_{\odot}$ and $T_{\mathrm{eff}}=2800\,$K, the escape velocity is about 36 km/s. Radial velocities derived from observations of the second overtone CO line ($\Delta v=3$) are of the order of 10-15 km/s. According to dynamical models, these lines are formed in the region from the stellar surface out to about 1.5 stellar radii \mbox{\citep{now10}}. Assuming an initial velocity of $u_{\mathrm{0}}=15$ km/s and a distance $R_0=1.5\,R_*$, Eq. (\ref{e_rmax}) gives a levitation distance of about $2\,R_*$. This corresponds approximately to where interferometric measurements place the inner edges of dust shells around AGB stars \citep[e.g.][]{witt07,kar11,norr12}.

Assuming a Planckian radiation field, geometrically diluted with distance from the star (i.e. an optically thin atmosphere with no molecular features), a power law for the dust  absorption coefficient $\kappa_{\mathrm{abs}}$ in the relevant wavelength range, and that the grain temperature is determined by the condition of radiative equilibrium, the condensation distance can be expressed as (see Appendix \ref{a_rcond})
\begin{equation}
\label{e_rcond}
\frac{R_\mathrm{c}}{R_*} = \frac{1}{2}\left(\frac{T_\mathrm{c}}{T_*}\right)^{-\frac{4+p}{2}}\quad\mathrm{where}\quad\kappa_{\mathrm{abs}}\propto \lambda^{-p},
\end{equation}
where $T_*$ is the effective temperature of the star. In Fig.~\ref{f_rcond} we have plotted curves of constant condensation distances as a function of $p$ and $T_{\mathrm{c}}$, using this formula. The chosen values of effective temperature, $T_*=2800\,$K and $T_*=2500\,$K, correspond to the detailed models used in this paper and by \citet{woi06fe}, respectively. Due to the wavelength dependence of the absorption coefficient (i.e. the value of $p$) different condensates will react differently to the close stellar environment. A grain material with a positive value of $p$ tends to heat up when interacting with the stellar radiation field, being more efficient at absorbing than emitting radiation, and thereby pushing the condensation distance further out. The opposite is true for a grain material with a negative value of $p$. To summarize, it is the combination of condensation temperature and the slope of the absorption coefficient that determines how close to a star a grain type can survive. For example, Fe-bearing silicates have a condensation temperature of about $1100\,$K and $p\approx 2.3$, which according to this simple formula results in a condensation distance of around 10$\,R_*$ for $T_*=2800$~K, far outside the reach of the levitated gas. The corresponding value for Fe-free silicates are $T_{\mathrm{c}}\approx1100\,$K and $p\approx-0.9$, resulting in a condensation distance of 2$\,R_*$ for $T_*=2800$~K, which is within reach of the levitated gas.

As mentioned in the introduction, phase-equilibrium cannot properly describe the dust formation in AGB-stars. According to detailed models, the formation process demands cooling well below the condensation temperature to initiate grain growth \citep[see, e.g.,][]{gailminerology}. Therefore, assuming thermal stability will lead to an underestimation of the condensation distance, and we treat it here as a lower limit. A 10\% decrease of the condensation temperature in Eq.~(\ref{e_rcond}) leads to an increase in condensation distance between \mbox{17-41\%}, depending on $p$, where a positive value of $p$ results in larger changes in $R_{\mathrm{c}}$. Despite the simplifications involved in Eq.~(\ref{e_rcond}), however, it turns out that the resulting condensation distances for amorphous carbon and Fe-free silicates compare well to RHD models which combine a detailed description of grain growth with a frequency-dependent treatment of radiative transfer \citep[e.g.][]{hof03,woi06fe,hof08bg}. These models place the condensation zone of such grains typically in the range 2-3$R_*$.

\subsection{Criteria for wind-drivers}
\label{s_grcrit}
To summarize, we have the following properties that determine if a specific grain material (pure or mixed) can be excluded as a wind-driver or be regarded as a potential candidate, where further investigation is needed:
\begin{description}
\item[ (a)]  The distance from the star where the grains are thermally stable, determined by the condensation temperature $T_{\mathrm{c}}$ and the near-infrared slope of the absorption coefficient $p$. 
\item[ (b)]  The abundance of the limiting element of the grain material.
\item[ (c)]  The absorption and/or scattering efficiency of the grain material in the wavelength region where most of the stellar radiation is emitted.
\end{description}
Further criteria, such as time-scales associated with growth rates that determine if a dust species can grow fast enough, have to be investigated with detailed non-equilibrium models of the atmospheric chemistry \citep[for a more in depth discussion on this topic see, e.g.,][or the references given in the introduction]{gailminerology}.

Estimates for specific grain materials can be obtained in the following way: if the particles can form sufficiently close to the stellar surface, as can be checked by Eq. (\ref{e_rmax}) and Eq.~(\ref{e_rcond}) (levitation distance and condensation distance, respectively), the combined effect of the critical abundance $\varepsilon_{\mathrm{lim}}$ and the efficiency $\qq$ can be investigated by setting $f_\mathrm{c}=1$ in the expression for the dust opacity $\kap$ (Eq. (\ref{e_kapspl})) and assuming a suitable flux distribution. If the resulting flux-averaged dust opacity $\kh$ is larger than the critical opacity $\kcrit$ the grain material is a possible candidate for triggering outflows. Examples of $\kh$ for different dust species, using the flux distribution plotted in the bottom panel of Fig. \ref{f_pl} (black curve), are given in Tab. \ref{t_gprop}. If one or more of the above criteria is not fulfilled for a specific grain material then we can exclude it as a wind-driver since the analytical approximations assume the most favorable limit.

\begin{figure}
\centering
\includegraphics[width=8.5cm]{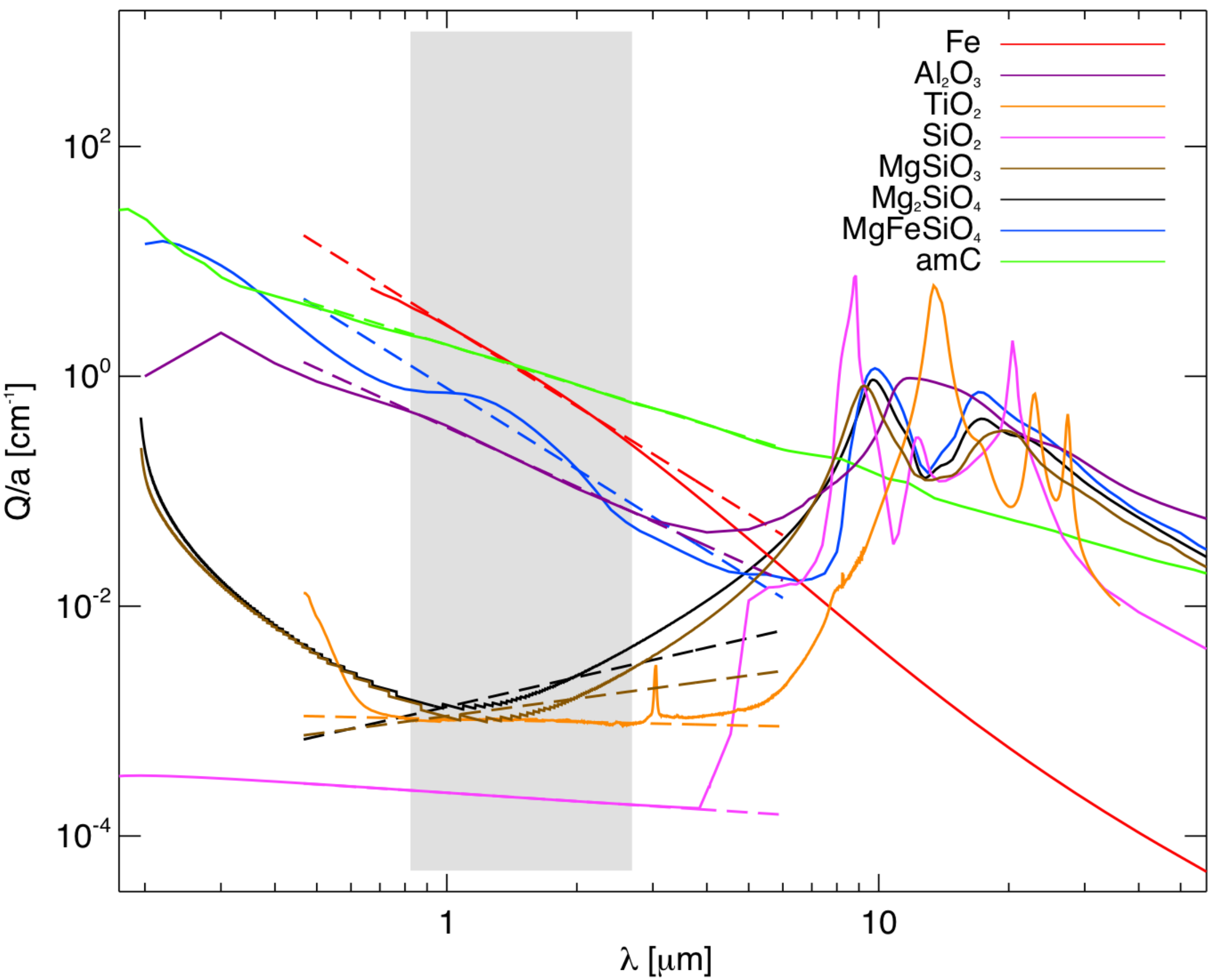}
\includegraphics[width=8.5cm]{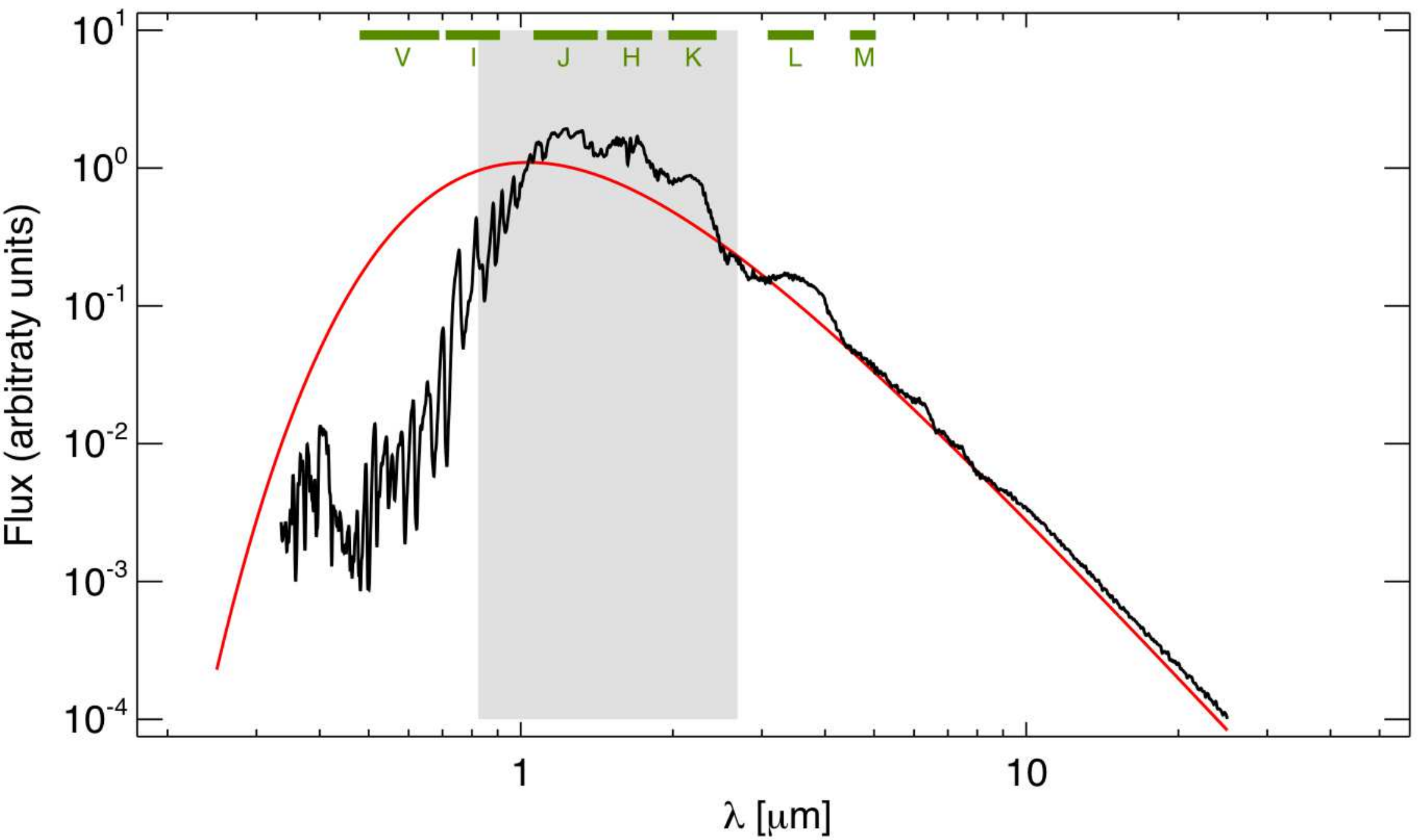}
   \caption{The top panel shows the efficiency per grain radius $\qq/\ag$, in the small particle limit, as a function of wavelength for a selection of dust species. The dashed lines show the power law fit according to \mbox{Eq. (\ref{e_pl})} and the shaded area indicates the wavelength region for which the optical data is fitted. The normalized flux of the hydrostatic initial model (black) and the Planck function for $T_{\mathrm{eff}}=2800\,$K (red) are plotted in the lower panel. Due to absorption by TiO molecules in the visual the flux of the initial model is very non-Planckian in this wavelength region.}
    \label{f_pl}
\end{figure}

\section{Detailed RHD models}
\label{s_model}
The analytical estimates discussed above presents a very simplified qualitative picture of the physics in the atmosphere of AGB stars; the dynamical effects of the gas pressure are ignored, the radiation field is described by a geometrically diluted Planck function and $\Gamma$ is considered as a simple function of the flux-averaged grain opacity. In reality, the molecular opacities strongly affect the radiation field, pulsation-induced shock waves influence the structure of the atmosphere and $\Gamma$ is a time-varying function depending on grain growth. To be able to take into account the complex interplay between gas dynamics and the radiation field we use state-of-the-art RHD models \citep[][and references therein]{hof03,hof08bg} that have been tested against different types of observations \citep[e.g.][]{gloidl04,now10,now11,sac11,bladh11}, and  can provide us with realistic temperature and density structures. In these numerical models we include a simple parameterized description of the dust opacity, designed to investigate what dust species can exist close to the stellar surface and potentially drive outflows.

\subsection{Gas dynamics and radiation field}
\label{s_dynmod}
The model atmosphere covers a spherical shell with an inner boundary situated just below the photosphere and an outer boundary in accordance with the dynamical properties of the model. In models that develop winds the outer boundary is fixed at the point where the flow velocity has reached its terminal value, allowing outflow, and in models without winds the outer boundary follows the periodic motions of the upper atmospheric layers. The variable structure of the atmosphere is described by the equations of hydrodynamics (equation of continuity, equation of motion and the energy equation) and the pulsations are simulated by temporal variation of physical quantities at the inner boundary. The opacities of molecules and dust forming in the outer cool layers of the atmosphere strongly affect the radiation field and in order to achieve realistic density and temperature structures the models include a frequency-dependent treatment of the radiative transfer \citep[see][for more details]{hof03}. 

The conservation of mass, momentum and energy is described by the following equations for the gas component:
\begin{eqnarray}
\label{e_hydro1}
&&\frac{\partial}{\partial t}(\rho)+\nabla\cdot(\rho u)=0\\
\label{e_hydro2}
&&\frac{\partial}{\partial t}(\rho u)+\nabla\cdot(\rho uu)=-\nabla P_{\mathrm{g}}-\frac{Gm_r}{r^2}\rho+\frac{4\pi\rho}{c}\left(\kappa^{\mathrm{g}}_{\mathrm{H}}+\kappa^{\mathrm{d}}_{\mathrm{H}}\right)H\\
\label{e_hydro3}
&&\frac{\partial}{\partial t}(\rho e)+\nabla\cdot(\rho eu)=-P_{\mathrm{g}}\nabla\cdot u+4\pi\rho\left(\kappa^{\mathrm{g}}_{\mathrm{J}}J-\kappa^{\mathrm{g}}_{\mathrm{S}}S_{\mathrm{g}}\right)
\end{eqnarray}
where $u$ is the radial velocity, $\rho$ is the gas density, $P_{\mathrm{g}}(\rho,e)$ is the thermal gas pressure, $m_r$ is the integrated mass within a spherical shell with radius $r$ and $e$ is the specific internal energy of the gas. We assume direct coupling between the motion of the gas and the dust, i.e. the momentum gained by the dust from the radiation field is directly transferred to the gas. Both the absorption and the scattering efficiency of the grain material may contribute to the overall momentum gain (see Eq. (\ref{e_qtot})) and both are included in $\kappa_{\mathrm{H}}^{\mathrm{d}}$. In Eq. (\ref{e_hydro2}) and (\ref{e_hydro3}) we use a simplified notation for the opacities compared to the rest of this paper. The superscripts $g$ and $d$ correspond to the gas and dust components respectively, and the subscripts indicate averages over different moments of the radiative intensity (see below). This means that $\kappa_{\mathrm{H}}^{\mathrm{d}}$ corresponds to $\kh$ as defined by Eq. (\ref{e_kh}), (\ref{e_kap2}) and (\ref{e_qtot}).
 
The energy budget of the dust component, and therefore the grain temperature, is determined by the condition of radiative equilibrium 
\begin{equation}
\label{e_radeq}
\kappa^{\mathrm{d}}_{\mathrm{abs,J}}J-\kappa^{\mathrm{d}}_{\mathrm{abs,S}}S(T_{\mathrm{d}})=0\longrightarrow T_{d}=\left(\frac{\kappa^{\mathrm{\mathrm{d}}}_{\mathrm{abs,J}}}{\kappa^{\mathrm{d}}_{\mathrm{abs,S}}}\right)^{1/4}T_{\mathrm{r}}.
\end{equation}
In this formula $T_r=\sqrt[4]{J\pi/\sigma}$ denotes the radiation temperature and $\kappa_{\mathrm{abs}}$ the true absorption part of the dust opacity. The frequency-integrated moments of the intensity, $J$ and $H$, are determined by solving the zeroth and first moment equation of the radiative transfer equation
\begin{eqnarray}
\label{e_radtrans1}
&&\nabla\cdot H+\rho{}\left(\kappa^{\mathrm{g}}_{\mathrm{J}}J-\kappa^{\mathrm{g}}_{\mathrm{S}}S_{\mathrm{g}}\right)=0\\
\label{e_radtrans2}
&&\nabla K+\frac{3K-J}{r}+\rho\left(\kappa^{\mathrm{g}}_{\mathrm{H}}+\kappa^{\mathrm{d}}_{\mathrm{H}}\right)H=0
\end{eqnarray}
simultaneously with Eq. (\ref{e_hydro1})-(\ref{e_hydro3}). Note that the terms containing the dust opacity cancel out in Eq. (\ref{e_radtrans1}) since we have assumed that radiative equilibrium holds for the dust grains (see Eq. (\ref{e_radeq})). 

The remaining unknown quantities required for closing the system of conservation laws (i.e. the frequency-averaged opacities, $\kappa_{\mathrm{J}}$, $\kappa_{\mathrm{H}}$ and $\kappa_{\mathrm{S}}$, and the Eddington factor $f_{\mathrm{edd}}=K/J$) are determined in a separate step: by solving the frequency-dependent radiative transfer equation for the current density-temperature structure after each hydrodynamical time-step we obtain $J_{\nu}$, $H_{\nu}$ and $K_{\nu}$, which allows us to compute the averaged opacities 
\begin{equation}
\label{e_opa}
\kappa_{\mathrm{X}}=\frac{1}{X}\int_0^{\infty} \kappa_{\nu}X_{\nu}\mathrm{d}\nu \quad\mathrm{where}\quad X=\int_0^{\infty}X_{\nu}d\nu.
\end{equation}
and the Eddington factor that are used in Eq. (\ref{e_hydro2})-(\ref{e_radtrans2}). In the current models we use 319 wavelength points, in contrast to 51 points in \citet{hof03}, for improved representation of the opacities and the radiation field. Assuming LTE, we approximate the source functions with Planck functions, i.e. $S_{\mathrm{g}}=B_{\nu}(T_{\mathrm{g}})$ and $S_{\mathrm{d}}=B_{\nu}(T_{\mathrm{d}})$, and the corresponding averaged opacities are Planck means.

It should be mentioned here that we neglect scattering when solving the frequency-dependent radiative transfer equation between hydro steps. Scattering on dust particles can have a significant effect on the momentum gain of individual grains and thus the wind dynamics. However, scattered photons will not noticeably change the overall radiation field, especially as the grains grow beyond the small particle limit and scattering becomes more forward oriented. Therefore the effect of scattering on the frequency-dependent intensity and its moments $J_{\nu}$, $H_{\nu}$, and $K_{\nu}$, which are required to compute the average opacities (see Eq. (\ref{e_opa})), will be small.\footnote{Note that scattering is included in the frequency-dependent opacities $\kappa_{\nu}$ in Eq.  (\ref{e_opa}) as appropriate for the respective means, i.e. $\kappa_{\nu}$ corresponds to $\kap$ when calculating the flux mean $\kh^{\mathrm{d}}$ that enters the equation of motion Eq. (\ref{e_hydro2}).} The exclusion of scattering in the radiative transfer leads to a significant reduction of the computational effort and allows us to use a larger number of frequency points, which is important for obtaining realistic temperature structures.

\subsection{Parameterized dust description}
\label{s_pardus}
\begin{figure}
\centering
\includegraphics[width=8.5cm]{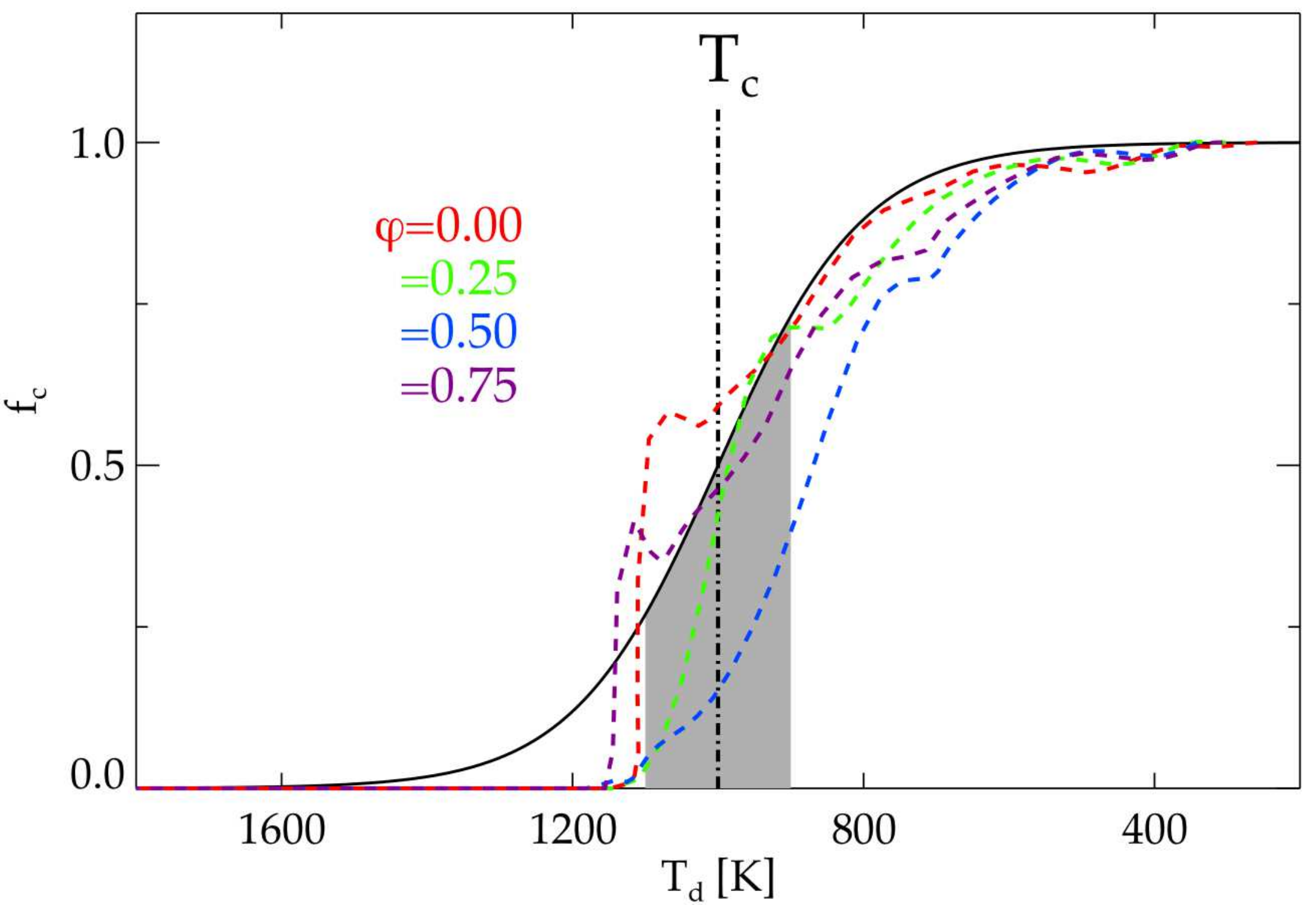}
   \caption{The degree of condensation as a function of grain temperature, using the parameterized dust description outlined in Sect.\ref{s_pardus} (solid black line), and for a model with a detailed description of Mg$_2$SiO$_4$ grains \citep[][model A; dashed lines]{hof08bg}. The different dashed curves represent a selection of pulsation phases $\varphi$ from the detailed model, where the degree of condensation has been normalized to vary between 0 and 1 to facilitate comparing the shape of the curves. The grey zone indicates how the parameter $\Delta T$ sets the width of the dust formation zone.}
      \label{f_fc}
\end{figure}

To systematically study the effect of a close stellar environment on different grain materials we construct a simple parameterized formula describing the dust opacity. This approach allows us to simulate aspects such as composite grain materials and effects of absorptions versus scattering cross-sections in a simple way, in addition to circumventing the problem of missing or incomplete optical data for certain dust species in the near infrared wavelength region \mbox{\citep[cf.][]{zeid11}}. Since we are interested in dust formation in the vicinity of a strong radiation source we focus on the factors in the expression for the grain opacity for which the radiation field will be important. One such factor is how the efficiency per grain radius, $\qq/\ag$, varies with wavelength, which in combination with the radiation field determines the grain temperature. Another important factor is the condensation temperature, a chemical property that indicates below which temperature grains of a certain material are thermally stable.  We therefore construct a parameterized dust description that allows for both different optical wavelength dependences and condensation temperatures, i.e.,
\begin{equation}
\label{e_parkap1}
\kap(\lambda) = \hat{\kappa}(\lambda)\cdot f_{\mathrm{c}}(r,t,T_{\mathrm{c}}).
\end{equation}
Our description is inspired by a formula used by \citet{b88} in his dust driven wind models, but is here generalized to allow for wavelength-dependent optical properties. The degree of condensation, $f_{\mathrm{c}}(r,t,T_{\mathrm{c}})$, is designed to increase monotonically with falling grain temperature, approaching a value of 1 as the grain temperature drops well below the condensation temperature (see Fig. \ref{f_fc}).
\begin{equation}
\label{e_fc}
f_{\mathrm{c}}(r,t,T_{\mathrm{c}}) = \frac{1}{1+e^{(T_{\mathrm{d}}(r,t)-T_{\mathrm{c}})/\Delta T}}
\end{equation}
The grain temperature $T_{\mathrm{d}}$ is determined by condition of radiative equilibrium (Eq. (\ref{e_radeq})) and it is a function of both time and distance from the star, due to the varying radiation field. The parameter $T_{\mathrm{c}}$ sets where $f_{\mathrm{c}}=0.5$ and $\Delta T$ regulates the width of the dust formation zone. The value of $\Delta T$ is chosen in accordance with models that include detailed dust formation (see Sect. \ref{s_mpid}). The wavelength-dependent part of the dust opacity $\hat{\kappa}$ is modeled as
\begin{equation}
 \label{e_pl}
\hat{\kappa}(\lambda) = \kappa_{\mathrm{0}}\left(\frac{\lambda}{\lambda_0}\right)^{-p}
\end{equation}
where $p$ is obtained by fitting a power-law function to $Q_{\mathrm{abs}}/\ag$ data in the wavelength region where most of the stellar flux is emitted (see Fig. \ref{f_pl} and Sect. \ref{s_mpid}). Regions of low flux will not contribute significantly to the energy balance and the radiative acceleration. In this expression $\kappa_{\mathrm{0}}$ is a scaling factor such that $\hat{\kappa}(\lambda_0)=\kappa_0$. In general the dust opacity would be a function of both wavelength and grain radius. However, since we are interested in the onset of grain formation and in particular the location of the dust formation zone, the adopted parameterization mimics the small particle limit where $\qq/\ag\approx Q_{\mathrm{abs}}/\ag$ is not a function of grain size (small particles have to exist before they can grow beyond the small particle limit).

To distinguish between effects of scattering and absorption of photons on dust grains we further introduce a quantity $f_{\mathrm{abs}}$ which sets the percentage of the dust opacity $\kap$ that is to be considered as true absorption
\begin{equation}
\label{e_parkap2}
\kappa_{\mathrm{abs}}(\lambda) = f_{\mathrm{abs}}\cdot\hat{\kappa}(\lambda)\cdot f_c(r,t)\quad\mathrm{where}\quad f_{\mathrm{abs}}=\frac{\kappa_{\mathrm{abs}}}{\kap}
\end{equation}
The dust opacity $\kap$, which includes contributions from both the absorption and scattering cross-sections, is used to calculate the radiative acceleration in the equation of motion, and the true absorption part $\kappa_{\mathrm{abs}}$ is used to determine the grain temperature (see Sect. \ref{s_dynmod}). This allows us to separate the dynamical and thermal effects of the dust opacity and the parameter $f_{\mathrm{abs}}$ can be adjusted to explore the effects of varying degrees of grain transparency. This parameter can also be used to simulate additional forces that affect the dynamics of the gas without changing the energy distribution of the radiation field.

\section{Model parameters}
\label{s_mpid}
\begin{figure*}
\centering
\includegraphics[width=18cm]{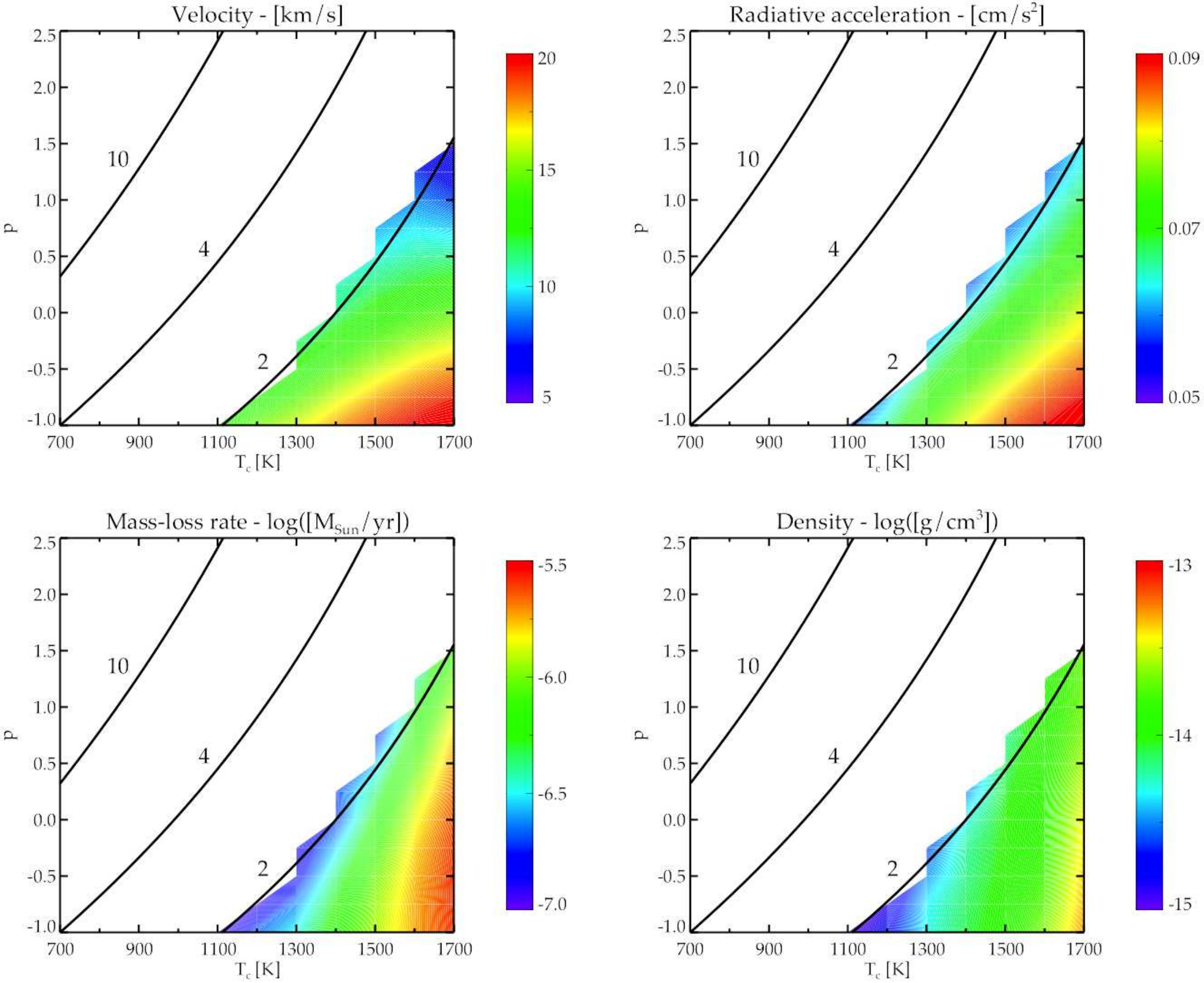}
   \caption{The panels show wind velocities, mass loss rates, radiative acceleration and density for a grid of dynamical models with stellar parameters $M_*=1\,M_{\odot}$, $L_*=5000\,L_{\odot}$ and $T_{\mathrm{eff}}=2800\,$K, a piston velocity of 4 km/s and $f_{\mathrm{abs}}=1.0$. The grid covers a range of chemical and optical dust properties, represented by the variables $T_{\mathrm{c}}$ and $p$, and includes in total 165 dynamical models. The sawtooth-like shape of the boundary between models with and without a wind is a consequence of the discrete number of models in the grid. The radiative acceleration and density is measured at the distance from the star when $f_{\mathrm{c}}=0.5$ (where $T_\mathrm{d}=T_\mathrm{c}$), averaged over pulsation cycle, and the velocity and mass loss rate are measured at the outer boundary of the model. The over-plotted contours are curves of constant condensation distance ($R_{\mathrm{c}}=2,4,10$) using \mbox{Eq. (\ref{e_rcond}).}}
   \label{f_ps10}
\end{figure*}

The stellar parameters chosen for the dynamical models in this study are typical for an M-type AGB star, with a stellar mass and luminosity of $1\,M_{\odot}$ and $5000\,L_{\odot}$, respectively, an effective temperature of $2800\,$K and solar abundances. The velocity amplitude used to simulate pulsations at the inner boundary was set to 4 km/s, which has proven to be a reasonable value when comparing with observations of molecular lines forming in the inner atmospheric region \citep{now10}. The same stellar parameters and piston velocity were used by \citet{hof08bg} in models with a detailed grain growth description for Mg$_2$SiO$_4$ particles, demonstrating that such grains can drive winds if they grow to sizes where scattering dominates the dust opacity.

Our formula for the dust opacity contains the optical parameters $p$ and $\kappa_0$, as well as the condensation temperature $T_{\mathrm{c}}$ and the width of the condensation zone $\Delta T$. For the purpose of this study $p$ and $T_{\mathrm{c}}$ will be considered as free parameters within a reasonable range. We focus on how the grain temperature, and consequently the condensation distance, is affected by the stellar radiation field (criterion (a) in Sect. \ref{s_grcrit}), disregarding the possibility of too low abundances or cross-sections for specific grain species (criteria (b) and (c)) in our grid of RHD models. We therefore choose a value for $\kappa_0$ such that $\Gamma>1$ when the grains have fully condensed ($f_{\mathrm{c}}=1$), making sure that the dust opacity will be high enough to trigger an outflow if grains condense sufficiently close to the star. For our chosen set of stellar parameters we find $\kcrit=2.6$ from Eq. (\ref{e_kcrit}). In the parameterized dust opacity we set $\kappa_{0}=3.0$ when $p=0$ and then adjust for other values of $p$ so that $\kh$ remains fixed. To provide good estimates for $\kappa_0$ we flux-average over the spectrum of the hydrostatic initial model instead of a Planckian radiation field,  since absorption by TiO molecules in the visual makes the flux distribution very non-Planckian. This results in a flux-averaged opacity $\kh\approx3.0$ and a corresponding value $\Gamma\approx1.1$, for all values of $p$ in the dynamical models. 

The parameter $\Delta T$ that adjusts the width of the dust formation zone in the parameterized dust description is set to $100\,$K. This value was chosen by comparing with the typical width of the dust formation zone in a  model with a detailed dust description and the same stellar parameters (see Fig. \ref{f_fc}). The exact value of $\Delta T$ does not have a significant effect on the resulting wind properties. 

Finally, with the stellar and dust parameters chosen, we set up a grid to investigate the dynamical effects of different combinations of chemical and optical properties in potential wind-driving dust species by varying the input parameters $p$ and $T_{\mathrm{c}}$ in the dust opacity. To cover the most probable combinations we let $p$ vary from -1 to 2.5 and $T_{\mathrm{c}}$ from $700\,$K up to $1700\,$K, in increments of 0.25 and $100\,$K respectively, including a total of 165 sampling points (individual dynamical models). In addition, we also vary the degree to which the dust opacity is considered true absorption by setting varying the parameter $f_{\mathrm{abs}}$  to 1.0 and 0.5 respectively (100\% and 50\% true absorption). 

\section{Results}
\label{s_res}
 \begin{figure}
\centering
\includegraphics[width=8.5cm]{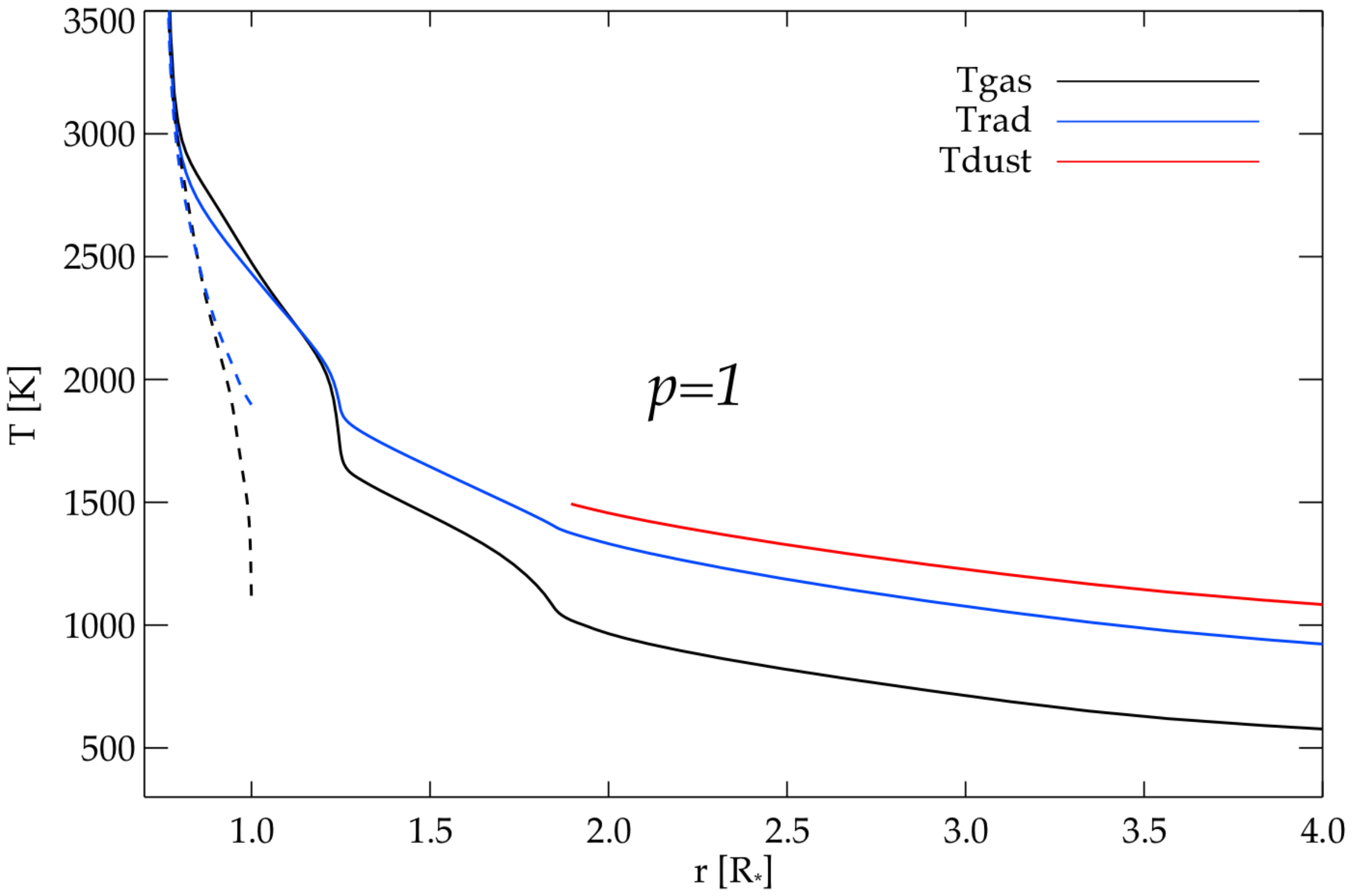}
\includegraphics[width=8.5cm]{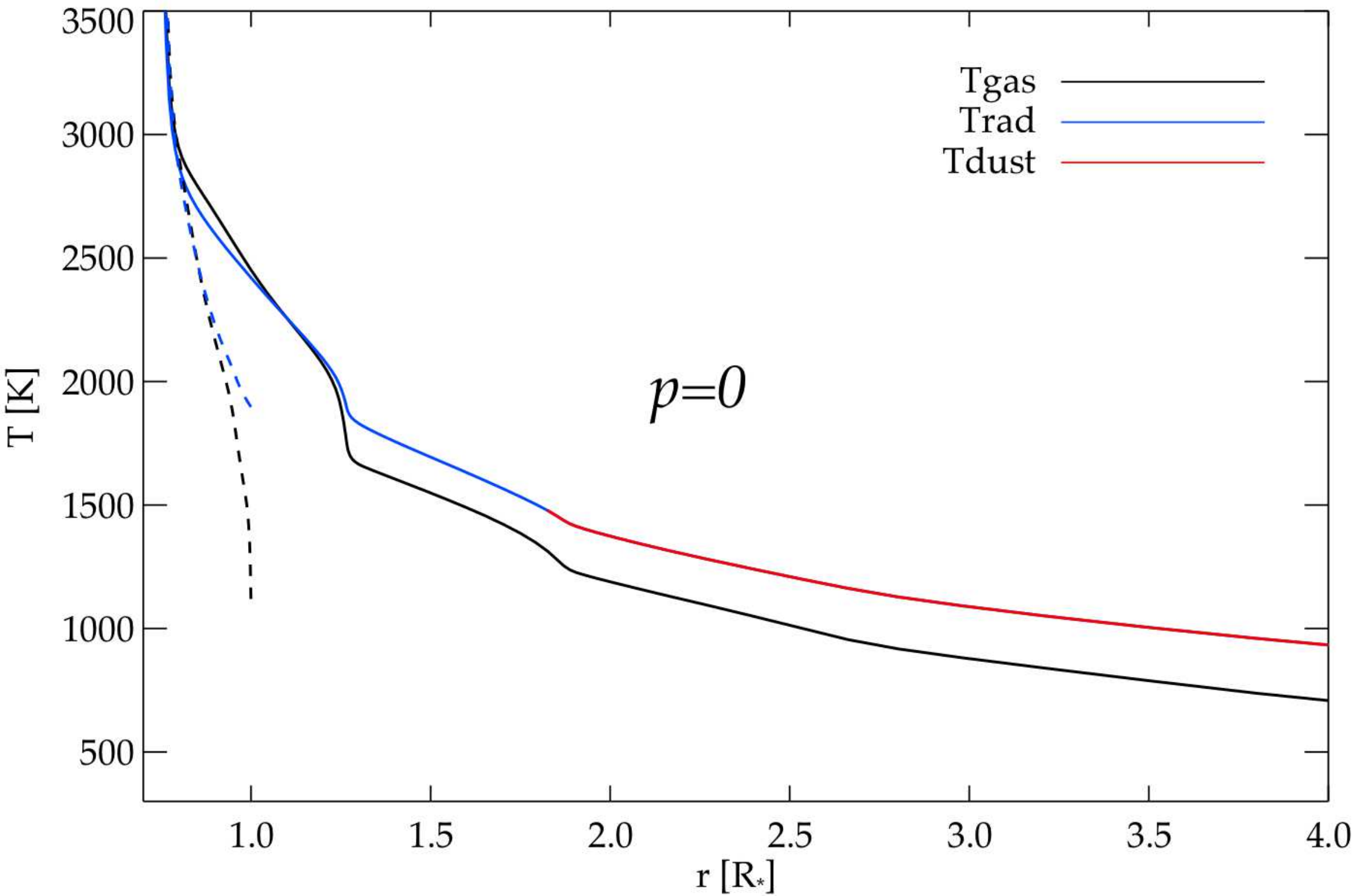}
\includegraphics[width=8.5cm]{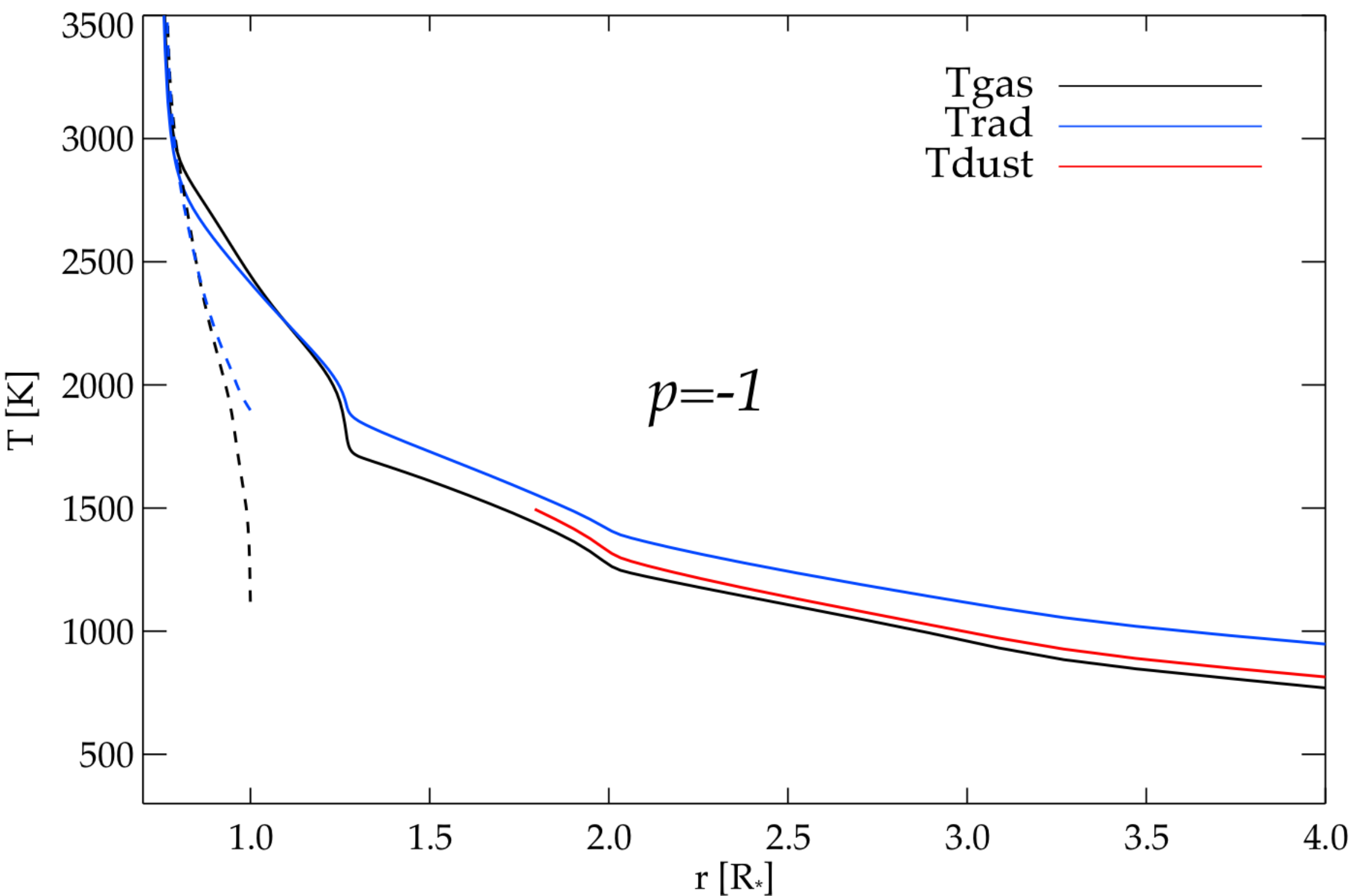}
   \caption{Dust-, gas- and radiation-temperatures as a function of distance for three different models where the condensation temperature and phase are kept fixed ($T_{\mathrm{c}}=1500\,$K and $\varphi=0.25$) but allowing for different values of $p$. The dashed lines show the temperature profiles of the hydrostatic starting model for the gas component (black) and the radiation field (blue).}
      \label{f_grtemp}
\end{figure}
\begin{figure}
\centering
\includegraphics[width=8.5cm]{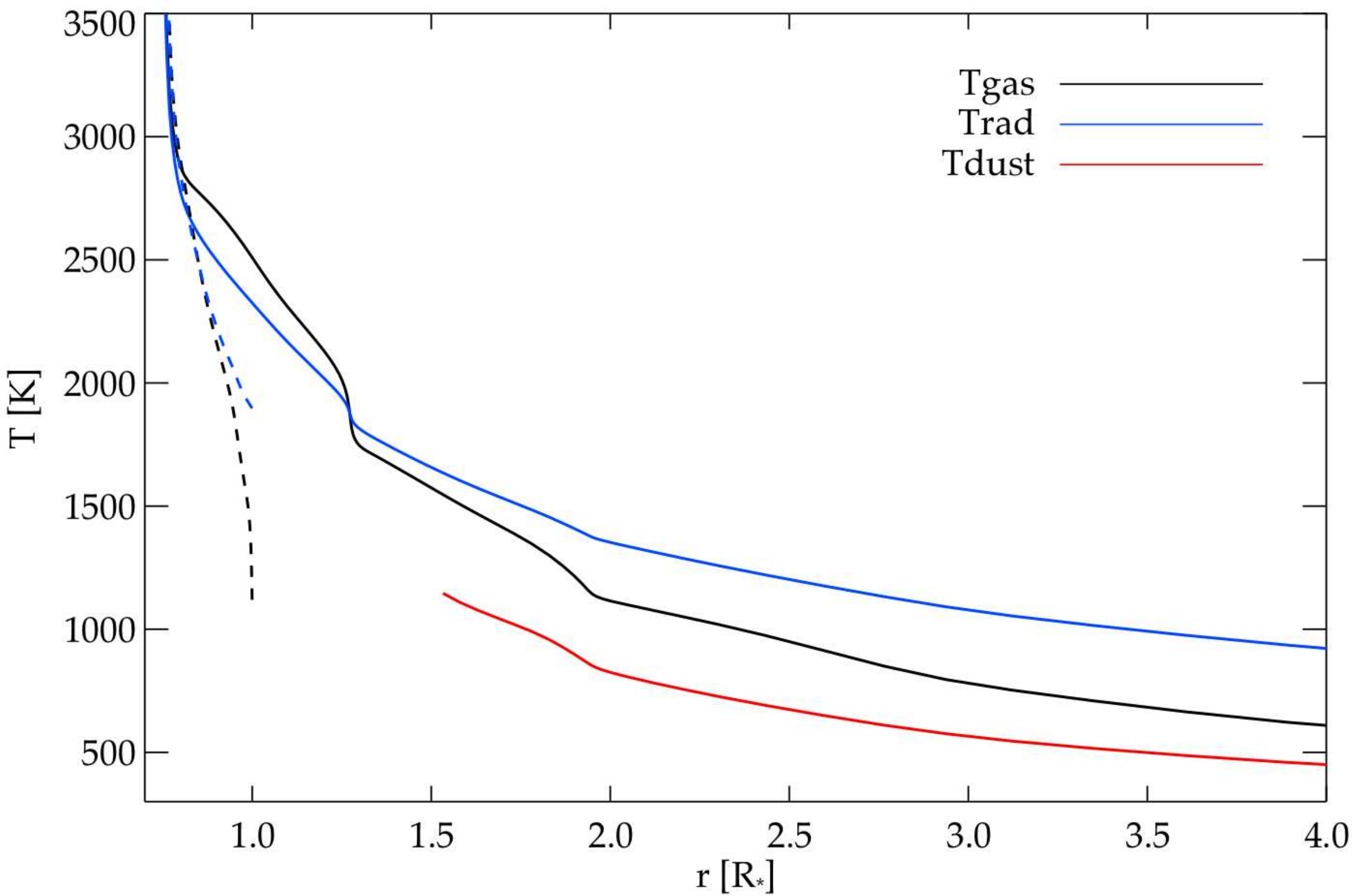}
   \caption{Dust-, gas- and radiation-temperatures as a function of distance for a model with a time-dependent description of the dust formation, using Mg$_2$SiO$_4$ grains \citep{hof08bg}. The dashed lines show the temperature profiles of the hydrostatic starting model for the gas component (black) and the radiation field (blue) at $\varphi=0.25$.} 
   \label{f_dmgrtemp}
\end{figure}
The detailed dynamical models provide us with the radial structure of the atmosphere and wind, from the photosphere to where the outflow reaches its terminal velocity \mbox{(typically $20-30\,R_*$)}, as a function of pulsation phase. This includes velocities, temperature- and density-structures and degree of condensation as a function of distance for the dust component. The mass loss rate and wind velocity at the outer boundary are recorded at each time-step and the values calculated for individual models are averaged over about 500 periods.

\subsection{Constraints on potential wind-drivers}
The grid of dynamical models shows very distinct trends as to what combination of optical and chemical dust properties, i.e. $T_{\mathrm{c}}$ and $p$, are necessary for driving a wind in M-type AGB stars (see Fig. \ref{f_ps10}). Dust material with high condensation temperatures and a NIR absorption coefficient that increases, or decreases slowly, with wavelength will condense close enough to the stellar surface to be able to trigger an outflow. As an example, dust materials with a condensation temperature around $1700\,$K and a negative slope as steep as $p\approx 1.5$ still produce winds. However, for condensation temperatures below $1400\,$K we need $p\leqslant 0$ to trigger an outflow.

The shape of the boundary between the wind and no-wind regime resembles the curves of constant condensation distance according to the simple estimate in Eq. (\ref{e_rcond}) (see the over-plotted contours in Fig. \ref{f_ps10}). This demonstrates the crucial role of the condensation distance in connecting the two stages of the mass loss scheme. While there is a qualitative agreement between the trends in the detailed models and the simple estimates it should be noted that the latter will not give quantitatively correct results. This can be seen in Fig. \ref{f_psrcond}, showing the actual condensation distance in the detailed RHD models.\footnote{Note in this context that the condensation distance is not a parameter in the detailed RHD models, but a result of the complex interplay between the dust properties and the radiation field.}

The exact location of the boundary between the wind and no-wind zone will depend on stellar parameters, pulsation amplitudes and certain physical assumptions made in the model. In particular, the simple parameterized dust opacity will tend to over-estimate the efficiency of dust formation due to the instant coupling between the degree of condensation and the grain temperature. That means that the dust opacity will take effect as soon as the grain temperature drops below the condensation temperature and the time-scales for grain growth are ignored. To compensate for this effect, we only include models that show well-developed winds, eliminating a few cases where a more sophisticated treatment of the dust formation probably would fail to produce stable outflows. Regarding stellar parameters, we focus the discussion in this paper on one set of values which are typical for M-type AGB stars, due to the computational efforts involved in creating a grid of this size. As far as the condensation distance is concerned, the most important stellar parameter is the effective temperature.  We can estimate the effect of a change in effective temperature by using Eq. (\ref{e_rcond}). As can be seen in Fig.~\ref{f_rcond}, the curves of constant condensation distance shift by about $100\,$K to left when we change $T_*=2800\,$K to $T_*=2500\,$K. The boundary between the wind and no-wind zones is therefore probably not as sharp as depicted in Fig.~\ref{f_ps10} and \ref{f_ps05}, but more like a smooth transition zone.

\subsection{Trends in grain temperature}
\label{s_grtemp}

\begin{figure*}
\centering
\includegraphics[width=18cm]{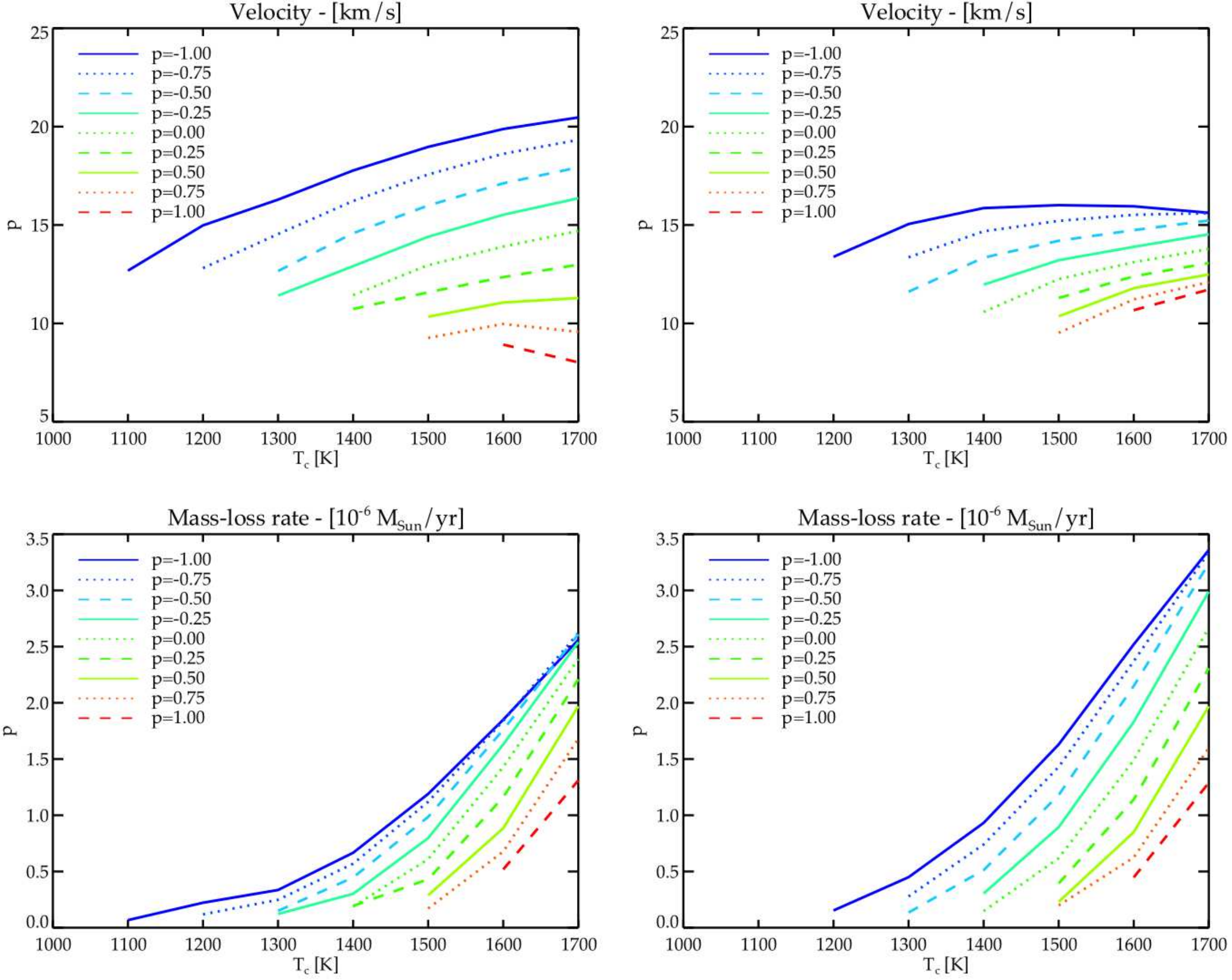}
   \caption{Trends in velocity and mass loss rate as a function of $T_{\mathrm{c}}$ for $p$ varying betwen $[-1,1]$. In the left two panels $f_{\mathrm{abs}}=1.0$ and in the right two panels $f_{\mathrm{abs}}=0.5$.}
   \label{f_trend}
\end{figure*}

The shape of the boundary between models with and without winds in the $p/T_{c}$-plane (see Fig. \ref{f_ps10}) is a consequence of the strong dependence of the grain temperature on $p$. A negative value of $p$ corresponds to a positive slope of the efficiency per grain radius $\qq/\ag$ (see Fig. \ref{f_pl}) in the wavelength region where most of the stellar flux is concentrated and materials with such optical properties will emit radiation more easily than they absorb. Therefore, in radiative equilibrium, the grain temperature will be lower than the radiation temperature, allowing the grains to form closer to the stellar surface. The opposite will happen for positive values of $p$, where the slope of $\qq/\ag$ is negative, and dust particles instead absorb radiation more easily than they emit, pushing the condensation distance further away from the stellar surface. This trend in grain temperature for positive versus negative values of $p$ is clearly visible in Fig. \ref{f_grtemp} where the temperature profiles for the gas component, the dust component and the radiation field are plotted as a function of distance for a fixed condensation temperature and phase, but with different values of $p$.  In the top panel $p=1$ and when dust starts to form the grain temperature becomes warmer than the radiation temperature. In the middle panel $p$ is set to zero and consequently the radiation temperature and dust temperature are equal. In the bottom panel $p=-1$ and we see a cooler dust temperature compared to the radiation temperature as soon as dust starts to form. 

For comparison we also plot the temperature structure for a model which includes a detailed description of Mg$_2$SiO$_4$ grains regarding both their formation and optical properties (see Fig. \ref{f_dmgrtemp}). Using the power law fit defined in Sect. \ref{s_pardus}, we obtain $p\approx-0.9$ for this material. However, as can be seen in Fig 3, a power law is not a good representation  of the data outside the marked grey area, underestimating the contribution to cooling at longer wavelengths. As a consequence, the grain temperature in the detailed model is even lower than for the corresponding parameterized case ($p=-1$, bottom panel of Fig. \ref{f_grtemp}).

\subsection{Trends in mass loss and wind velocity}

Since we chose $\kappa_0$ to produce a flux-averaged dust opacity large enough to trigger a wind when grains have fully condensed ($\Gamma>1$ for $f_{c}=1$), dynamical properties like mass loss rates and wind velocities will not be representative of specific dust species. Despite this, there is still interesting information to derive from the overall trends in mass loss rate and wind velocity in this parameter study.  For the mass loss rates, as can be seen in the bottom panels of Fig. \ref{f_trend}, there is a strong correlation between increasing condensation temperature and increasing mass loss rates. There is also a weaker correlation between decreasing value of the power law coefficient $p$ and increasing mass loss rates. Both increasing the condensation temperature and decreasing the value of the power law coefficient pushes the condensation distance closer to the stellar surface, where the density $\rho$ is higher, which is what causes the higher mass loss rates since $\dot{M}\propto \rho$. The correlation between mass loss rate and density at the condensation distance is evident in the bottom panels of Fig. \ref{f_ps10}.

The radiative acceleration, and consequently the velocity, of the out-flowing gas depends on the flux-integrated dust opacity $\kh$. This quantity is affected by both the spectral energy distribution of the stellar flux and the power law representation of the dust opacity  (see Eq. (\ref{e_kh})). As mentioned above, increasing the condensation temperature and decreasing the value of the power law coefficient pushes the condensation distance closer to the stellar surface and results in higher mass loss rates. As dust starts to form and the optical depth of the wind increases, the spectral distribution of the radiation field will shift towards longer wavelengths compared to the photospheric energy distribution. The redding of the radiation field will be more pronounced the higher the mass loss rate is and the more opaque the dust particles are. 

The radiative acceleration of dust material with different power law coefficients will be affected differently by the reddening of the radiation field: a negative value of $p$ (positive slope of $Q/a_{\mathrm{gr}}$) will result in a higher radiative acceleration compared to a positive value of $p$ (negative slope of $Q/a_{\mathrm{gr}}$), due to the increased radiation in the far infrared. This gives rise an increasing spread in velocity for different values of $p$ with increasing condensation temperature, which can be seen in the top left panel in Fig. \ref{f_trend}. This trend is less pronounced (see the top right panel in Fig. \ref{f_trend}) when we assume more transparent grains by decreasing the value of $f_{\mathrm{abs}}$ without decreasing the radiative acceleration, since the reddening of the radiation field is reduced when the dust absorption is lowered.

\subsection{Effects of varying degree of grain transparency}
\begin{figure*}
\centering
\includegraphics[width=18cm]{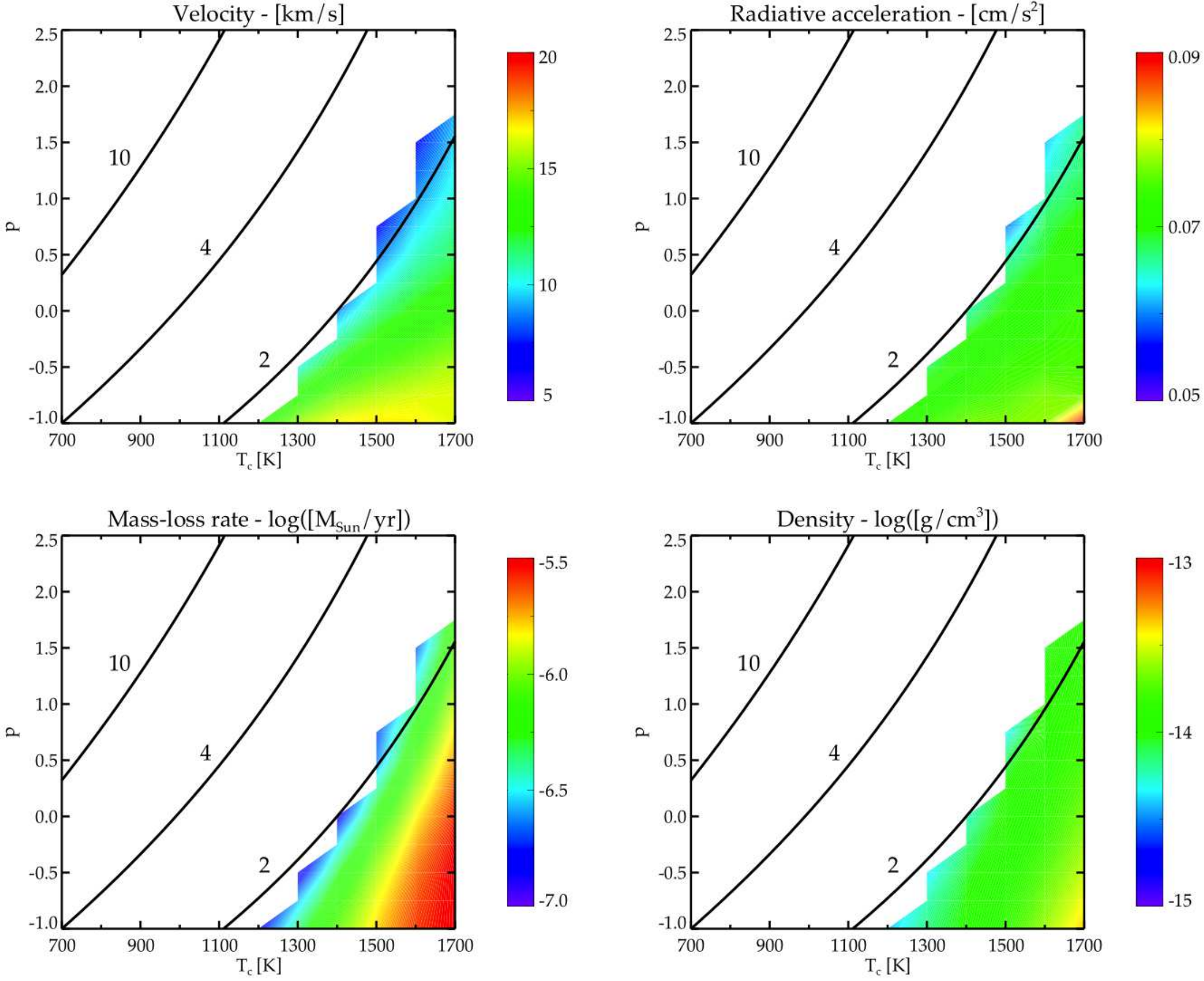}
   \caption{Same as Fig. \ref{f_ps10}, but for $f_{\mathrm{abs}}=0.5$.}
   \label{f_ps05}
\end{figure*}
\begin{figure}
\centering
\includegraphics[width=8.5cm]{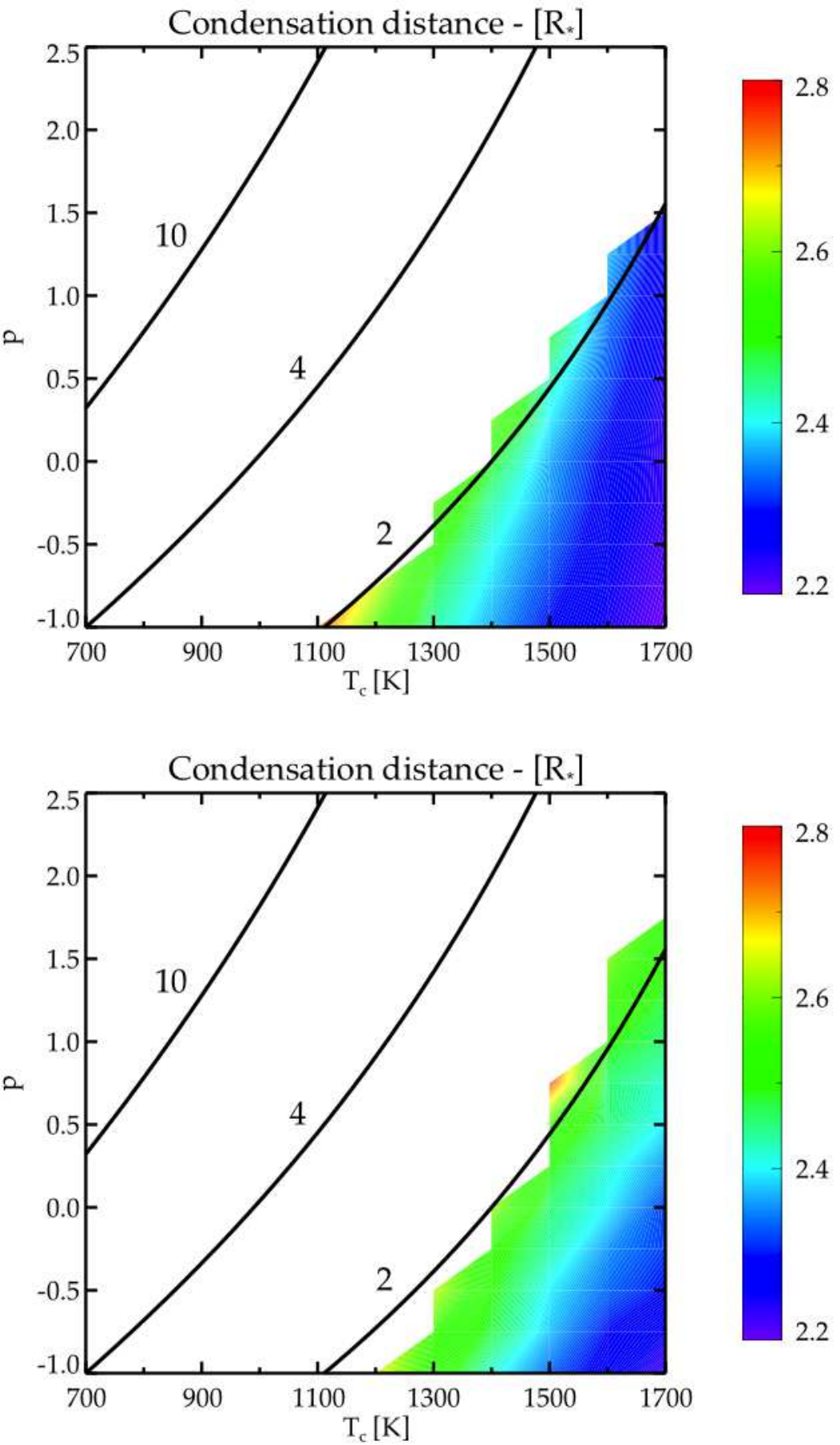}
   \caption{Distance from the stellar surface where the degree of condensation has reached $f_{\mathrm{c}}=0.5$ (where $T_\mathrm{d}=T_\mathrm{c}$), averaged over pulsation cycle. In the top panel $f_{\mathrm{abs}}=1.0$ and in the bottom panel $f_{\mathrm{abs}}=0.5$.}
   \label{f_psrcond}
\end{figure}

In addition to exploring the dynamical effects of different optical wavelength dependences and condensation temperatures we also investigate the effects of grains that show various degrees of transparency. Both scattering and true absorption may contribute to the radiative acceleration of dust particles (see Sect.\ref{s_radacc}) and scattering might even be the dominant process for momentum transfer, as demonstrated for Mg$_2$SiO$_4$ grains by \citet{hof08bg}. As in the case of the different values of $p$, we keep the radiative acceleration fixed but vary the fraction of $\kap$ that is considered as true absorption. For the values of $f_{\mathrm{abs}}$ that we have tested we obtain a similar pattern in the parameter space concerning the wind/no-wind zone; combinations of high positive $p$-values and low condensation temperatures produce no outflows (see the right panels of Fig. \ref{f_ps10} and \ref{f_ps05}). The location of the boundary between the wind and the no-wind regime is, however, slightly shifted. This is a consequence of how the dust opacity affects the flux distribution. A higher degree of true absorption by the dust grains results in a stronger reddening and thermalization of the radiation field. The larger optical depth of the circumstellar envelope leads to a more Planckian radiation field. This may explain why the shape of the boundary in Fig. \ref{f_ps10} follows the contours of constant $R_{\mathrm{c}}$ derived from Eq. (\ref{e_rcond}), which is based on a Planckian flux distribution, more closely.

A detailed comparison between the models with $f_{\mathrm{abs}}=0.5$ and $f_{\mathrm{abs}}=1.0$ shows somewhat different dynamical structures in the atmospheres that produce winds. The mass loss rate is strongly correlated with the density in the dust formation zone in both cases (see the bottom panels of Fig. \ref{f_ps10} and \ref{f_ps05}), but the radial position of the zone itself varies less for models with $f_{\mathrm{abs}}=0.5$ (see Tab. \ref{t_dyn}). The smaller variation in velocity in the grid with $f_{\mathrm{abs}}$ set to 0.5 is expected, since the circumstellar reddening of the radiation field is less pronounced when dust particles absorb less of the stellar flux. The variation in mass loss rates for the two sets of models is a consequence of the variation in the density at the condensation distance.

\begin{table}
\caption{Ranges of dynamical and atmospheric properties in models producing outflows in the two RHD grids ($f_{\mathrm{abs}}=0.5$ and $f_{\mathrm{abs}}=1.0$).}             
\label{t_dyn}      
\centering                          
\begin{tabular}{l l c c}        
\hline\hline                 
 & & $f_{\mathrm{abs}}=0.5$ & $f_{\mathrm{abs}}=1.0$ \\
\hline       
$u$ & [km/s] & $9.5-16.0$ & $5.9-20.5$\\
$\dot{M}$ & [$M_{\odot}$/yr] & $1\cdot 10^{-7}-3\cdot10^{-6}$ & $7\cdot 10^{-8}-3\cdot10^{-6}$ \\
$R_{\mathrm{c}}$ & [$R_*$] & $2.3-2.5$ & $2.2-2.8$ \\
$\rho(T_{\mathrm{c}})$ & [g/cm$^2$] & $4\cdot10^{-15}-4\cdot 10^{-14}$ & $6\cdot10^{-16}-5\cdot 10^{-14}$ \\
\hline   
\end{tabular}
\end{table}

\section{Discussion: specific grain materials}
\label{s_spg}

To demonstrate how the result presented in the previous sections can be used to evaluate potential wind-drivers, let us take a closer look at a few specific grain materials relevant for M-type AGB stars. Typical values of $p$ can be derived from optical data of actual grain materials, as shown in Fig. \ref{f_pl} and Tab. \ref{t_gprop}.\footnote{The wavelength region where the power law is fitted, indicated by the grey area in Fig. \ref{f_pl}, corresponds to where the condition $E>|0.5E_{\mathrm{max}}|$ is satisfied. In this expression $E=\lambda B_{\lambda}(T_*,\lambda)$, where $E_{\mathrm{max}}=E(\lambda_{max})$ and $\lambda_{max}$ is given by Wien's law.} It is clear that metallic iron, Fe-bearing silicates and amorphous carbon can all be fitted well by a power-law function, even quite a bit outside the wavelength range where the star radiates most of its flux. A power-law fit for the Fe-free silicate, however, is less appropriate outside the considered wavelength range and while this approach captures the essential dynamical effects of the dust material, care has to be taken when considering photometric fluxes outside the fitted region \mbox{\citep[][in prep.]{bladhip}.}

The presence of metallic iron in the circumstellar environment of AGB stars has been suggested based on observations \citep[e.g][]{ian10}. This material is very opaque in the near-infrared wavelength region (see Fig. \ref{f_pl}). However, assuming that all available iron has condensed into dust and averaging over the photospheric flux distribution results in a total opacity which is lower than the critical opacity needed to drive a wind. More importantly, due to the combination of $p\approx2.4$ and $T_{\mathrm{c}}\approx1050\,$K for this material, iron grains cannot form at distances that can be reached by the shock-levitated gas and can therefore not trigger an outflow (see Fig.  \ref{f_rcond} and  \ref{f_ps10}).

A better suggestion for possible wind-drivers are Fe-bearing silicates (MgFeSiO$_4$). They have an absorption efficiency comparable to metallic iron in the near-infrared but are not as compact (lower density). This results in a flux-averaged dust opacity above the critical opacity for our chosen stellar parameters, when assuming full condensation. But as can be seen from Fig. \ref{f_rcond} and \ref{f_ps10}, the optical and chemical properties of Fe-bearing silicates ($p\approx2.3$ and $T_{\mathrm{c}}\approx1100\,$K) prevent them from forming sufficiently close to the stellar surface. 

In contrast, Fe-free silicates such as Mg$_2$SiO$_4$ or MgSiO$_3$ can form in close proximity to the stellar surface but are very transparent in the wavelength range around 1~$\mu$m, leading to a total opacity well below the critical opacity as long at the particles are small. But if the particles grow to grain sizes comparable to the wavelength of the flux maximum, the contribution to the efficiency $\qq$ from the scattering cross-section is substantial, and for the case of Mg$_2$SiO$_4$ grains, scattering has been demonstrated to be sufficient to drive a wind \citep[see][Fig.~1]{hof08bg}. In the framework of our parameterized RHD models, this material has  $p\approx-0.9$ and $T_{\mathrm{c}}\approx1100\,$K, which puts it right on the border of the wind/no-wind zone in Fig. \ref{f_ps10}. However, as discussed in Sect. \ref{s_grtemp}, the power law fit tends to overestimate the grain temperature compared to the real optical data, resulting in a larger condensation distance than in a detailed model.

The transparency of TiO$_2$ together with the low abundance of the limiting element rules this material out as a potential wind-driver, even if it can form close to the stellar surface. Lastly, it is difficult to say anything definite about dust species like SiO$_2$ or Al$_2$O$_3$ at present due to uncertainties in measured optical data and incomplete wavelength coverage. 

\begin{table*}
\caption{Compilation of properties for a few selected dust species.}             
\label{t_gcrit}      
\centering                          
\begin{tabular}{l c c c c c}        
\hline\hline                 
Grain material &Condensation & Abundance (b) & \multicolumn{2}{c}{Cross-section (c)}  & Wind-driver\\    
 & distance (a) & & SPL & BG & \\
\hline       
   Fe &  NO & YES & YES & $-$ & NO\\
   Al$_2$O$_3$ & YES & NO & NO & $-$ & NO\\
   TiO$_2$ & YES & NO & NO & $-$ & NO\\
   SiO$_2$ & ? & YES & NO & ?  & ?\\   
   MgSiO$_3$ & YES & YES & NO & ? & ?\\
   Mg$_2$SiO$_4$ & YES & YES & NO & YES & YES\\
   MgFeSiO$_4$ &  NO & YES & YES & $-$ & NO\\ 
\hline
   amC & YES & YES & YES & YES & YES\\   
\hline                                   
\end{tabular}
\tablefoot{A compilation of the criteria mentioned in Sec \ref{s_grcrit} for the dust species listed in Tab. \ref{t_gprop}. The question mark in condensation distance and cross-section for SiO$_2$ and Al$_2$O$_3$ is due to uncertainties in  the optical data. Note that amC is a wind-driver in C-type AGB stars whereas all other dust species are considered in the context of M-type AGB stars. For figures showing dust opacities beyond the small particle limit, see \citet{hof08bg} and \citet{matt11} for Mg$_2$SiO$_4$ and amC, respectively.}
\end{table*}

Optical properties are predominantly measured in wavelength regions where there are distinct spectral features and laboratory studies often leave out regions where opacities are expected to be comparatively low \citep[see discussion in][]{zeid11}. Characteristic features are important for identifying individual grain materials but may not be essential for the overall energy balance of the circumstellar environment. In contrast, the optical properties in regions of low absorption may be crucial for grain temperatures and radiative acceleration, if they coincide with regions of high stellar flux.

As an example, consider the optical data for Al$_2$O$_3$ provided by \cite{koi95cor} and the extrapolated data used by \cite{woi06fe}. The slopes of the absorption coefficients in the near-infrared differ dramatically for these two sets of data, with $p\approx 1.7$ and  $p\approx -2.1$, respectively. The resulting grain temperatures at a given distance from the star will therefore be very different, affecting the condensation distance. The numerical results in \cite{woi06fe} suggests that Al$_2$O$_3$ grains form as close as 1.5~$R_*$ to the stellar surface, whereas a simple estimate for the data from  \cite{koi95cor}, using Eq. (\ref{e_rcond}), results in a condensation distance of about 3~$R_*$ for $T_*=2500$~K. However, setting aside uncertainties in optical data and the condensation distance, Al$_2$O$_3$  is an unlikely candidate since the elemental abundance of Al is 1-2 orders of magnitude lower than for most of the limiting elements in Tab. \ref{t_gprop}.

It is likely that several different kinds of dust species will form close to the stellar photosphere and not all of them will fulfill the criteria concerning abundance and cross-section that are required to trigger outflows. However, some of them may act as seed particles for the actual wind-drivers. We might also expect silicate grains to be enriched by iron further out in the wind, even if such grains cannot form close to the stellar surface.

To summarize the above discussion we compile the criteria for wind-drivers in Tab. \ref{t_gcrit}. The criterion concerning the cross-section has been divided into two columns, one for the small particle limit (SPL) and one for bigger grains where scattering may play a role (BG). The column with cross-sections for grain sizes beyond the small particle limit has been filled in when the contribution by scattering may be the decisive factor for making a material a potential wind-driver (as in the case of SiO$_2$, MgSiO$_3$ and Mg$_2$SiO$_4$). The chemical and optical properties of Al$_2$O$_3$, TiO$_2$ and MgSiO$_3$ place them just outside the part of the parameter space producing outflows (see Fig. \ref{f_rcond} and \ref{f_ps10}). However, they have been listed as satisfying the condition of condensation distance, since the boundary between the wind/no-wind zones might be slightly shifted due to varying pulsation amplitudes and stellar parameters (see Fig. \ref{f_rcond}). More examples of $p$ and $T_{\mathrm{c}}$ values for dust species that are not considered potential wind-drivers, due to low abundances or large condensation distances, are given in Tab. \ref{t_more}.

We have included amorphous carbon in Tab. \ref{t_gprop} and \ref{t_gcrit} for comparison because it is a known wind-driver for C-type AGB stars. Amorphous carbon (amC) is very opaque in the relevant wavelength region, it has a high condensation temperature and the limiting element is abundant in carbon-rich atmospheres. There is, however, practically no carbon available for forming dust particles in M-type AGB stars, since most carbon atoms are locked in the tightly bound CO-molecule. The optical and chemical properties of this dust species ($p\approx1.2$ and $T_{\mathrm{c}}\approx1700\,$K) place it in a different region of the parameter space than the grain materials discussed above. Therefore amC grains will affect the spectra in a different way than e.g. Fe-free silicate particles, as will be discussed in a forthcoming paper \citep{bladhip}.

\begin{table}
\caption{More examples of $p$ and $T_{\mathrm{c}}$ values (column 2 and 3) for dust species in the circumstellar environments of AGB stars, as inferred from observations \citep{agbgrain}. In column 4 the condensation distances $R_{\mathrm{c}}$, using Eq. (\ref{e_rcond}), are listed.}             
\label{t_more}      
\centering                          
\begin{tabular}{l r c c l}        
\hline\hline                 
Material & $p$ & $T_{\mathrm{c}}$ [K]  & $R_{\mathrm{c}}/R_*$ &Ref. ($T_{\mathrm{c}}$ and $p$)\\    
\hline       
   MgAl$_2$O$_4$ & $-1.2$ & 1150 & 1.7 & 1, 3. \\
   Mg$_{0.5}$Fe$_{0.5}$SiO$_3$ & 2.4 & 1100 & 9.9 & 1, 2\\	
   Fe$_2$SiO$_4$ & 1.7 & 1050 & 8.2 & 1, 3\\
   Mg$_{0.5}$Fe$_{0.5}$O & 1.9 & 1000 & 10.4 & 1, 2.\\
   FeO & 1.6 & 900 & 12.0 & 1, 2.\\
\hline   
   SiC & $0.3$ & 1100 & 3.7 & 1, 4. C-rich atmos.\\
\hline   
\end{tabular}
\tablebib{(1) \citet{gailminerology}; (2) \citet{hen95}; (3) \citet{zeid11}; (4) \citet{peg88sic}.
}
\end{table}

\section{Summary and conclusions}
\label{s_concl}
Observations of molecular lines in the inner part of the atmosphere provide constraints on how far shock waves can levitate gas (i.e. a few stellar radii). A necessary condition for the occurrence of  dust-driven outflows is that the condensation distance falls within this range. However, it is not sufficient that grains of a certain material can form within reach of the levitated gas in order to trigger a wind (criterion (a) in Sec \ref{s_grcrit}). The individual dust species also need radiative cross-sections and abundances such that they can provide the radiative acceleration necessary to cause a general outflow (criteria (b) and (c) in Sec \ref{s_grcrit}).

To investigate how a close stellar environment affects dust formation we have a constructed a grid of detailed atmosphere and wind models, including a frequency-dependent treatment of radiative transfer for the gas and dust components. We see distinct trends in the numerical results concerning what combinations of optical and chemical dust properties (here quantified by the parameters $T_{\mathrm{c}}$ and $p$) are required to trigger an outflow. Dust species with a low condensation temperature and a near-infrared absorption coefficient that decreases strongly with wavelength will not condense close enough to the stellar surface to be considered as potential wind-drivers. The grain temperature, and consequently the condensation distance, is strongly influenced by the wavelength dependence of the absorption coefficient. A simple estimate (see Eq. (\ref{e_rcond})) of how the interaction between a non-grey dust opacity and the stellar radiation field affects the condensation distance shows good qualitative agreement with the detailed numerical models. However, the quantitive results differ due to the inherent physical simplifications in the analytical estimate.

A frequency-dependent treatment of the dust opacity is absolutely necessary when calculating dust yields from AGB stars. A grey treatment of the dust opacity (corresponding to $p=0$) will underestimate the condensation distance for dust species with a power law coefficient $p>0$ while the opposite is true for dust species where $p<0$, producing misleading results regarding the overall dust composition. 

As for specific grain materials, the wavelength dependence of the near-infrared absorption coefficients for metallic iron and Fe-bearing silicates, together with their low condensation temperatures, prevent them from forming close to the stellar surface, in accordance with earlier results by \citet{woi06fe}. This means that neither metallic iron or Fe-bearing silicates are viable wind-drivers. TiO$_2$ is also excluded as a candidate, but in this case it is a consequence of the low abundance of the limiting element, Ti, and the low cross-section. Other cases, such a SiO$_2$ and Al$_2$O$_3$, are less clear-cut due to uncertainties in the dust data and further work is needed. In particular, precise optical data in the near-infrared where AGB stars emit most of their flux is required for a reliable determination of grain temperatures. 

A dust species that has been shown to work as a wind-driver if the grains grow to sizes around 0.1-1 micron is Mg$_2$SiO$_4$ \citep{hof08bg}. It consists of relatively abundant elements and the condensation temperature together with the wavelength-dependent dust opacity allows it to form close to the stellar surface ($\sim2\,R_*$). These grains are very transparent in the near-infrared in the small particle limit, but if the grains reach sizes comparable to the wavelength of the stellar flux maximum and scattering is taken into account, the overall cross-section can provide enough radiative acceleration to trigger a wind. This scenario is supported by recent observations of grains in the critical size-range (i.e. $\sim0.3\,\mu$m) in the close stellar environment \citep{norr12}.

\begin{acknowledgements}
      This work has been supported by the Swedish Research Council (\textit{Vetenskapsr\aa det}). The computations were performed on resources provided by the Swedish National Infrastructure for Computing (SNIC) at UPPMAX.
\end{acknowledgements}

\bibliographystyle{aa}
\citeindextrue
\bibliography{references}

\begin{thebibliography}{46}
\expandafter\ifx\csname natexlab\endcsname\relax\def\natexlab#1{#1}\fi

\bibitem[{{Andersen} {et~al.}(2003){Andersen}, {H{\"o}fner}, \&
  {Gautschy-Loidl}}]{and03}
{Andersen}, A.~C., {H{\"o}fner}, S., \& {Gautschy-Loidl}, R. 2003, \aap, 400,
  981

\bibitem[{{Begemann} {et~al.}(1997){Begemann}, {Dorschner}, {Henning},
  {Mutschke}, {Guertler}, {Koempe}, \& {Nass}}]{beg97}
{Begemann}, B., {Dorschner}, J., {Henning}, T., {et~al.} 1997, \apj, 476, 199

\bibitem[{{Bladh} {et~al.}(2011){Bladh}, {H{\"o}fner}, \& {Aringer}}]{bladh11}
{Bladh}, S., {H{\"o}fner}, S., \& {Aringer}, B. 2011, in Astronomical Society
  of the Pacific Conference Series, Vol. 445, Why Galaxies Care about AGB Stars
  II: Shining Examples and Common Inhabitants, ed. {F.~Kerschbaum,
  T.~Lebzelter, \& R.~F.~Wing}, 315

\bibitem[{{Bladh} {et~al.}(2012){Bladh}, {H{\"o}fner}, {Nowotny}, \&
  {Aringer}}]{bladhip}
{Bladh}, S., {H{\"o}fner}, S., {Nowotny}, W., \& {Aringer}, B. 2012

\bibitem[{{Bohren} \& {Huffman}(1983)}]{bohuff83}
{Bohren}, C.~F. \& {Huffman}, D.~R. 1983, {Absorption and scattering of light
  by small particles}, ed. {Bohren, C.~F.~\& Huffman, D.~R.}

\bibitem[{{Bowen}(1988)}]{b88}
{Bowen}, G.~H. 1988, \apj, 329, 299

\bibitem[{{Dorschner} {et~al.}(1995){Dorschner}, {Begemann}, {Henning},
  {Jaeger}, \& {Mutschke}}]{dor95}
{Dorschner}, J., {Begemann}, B., {Henning}, T., {Jaeger}, C., \& {Mutschke}, H.
  1995, \aap, 300, 503

\bibitem[{{Ferrarotti} \& {Gail}(2001)}]{fega01}
{Ferrarotti}, A.~S. \& {Gail}, H.-P. 2001, \aap, 371, 133

\bibitem[{{Ferrarotti} \& {Gail}(2006)}]{fega06}
{Ferrarotti}, A.~S. \& {Gail}, H.-P. 2006, \aap, 447, 553

\bibitem[{{Fleischer} {et~al.}(1992){Fleischer}, {Gauger}, \&
  {Sedlmayr}}]{flei92}
{Fleischer}, A.~J., {Gauger}, A., \& {Sedlmayr}, E. 1992, \aap, 266, 321

\bibitem[{{Gail}(2010)}]{gailminerology}
{Gail}, H.-P. 2010, in Lecture Notes in Physics, Berlin Springer Verlag, Vol.
  815, Lecture Notes in Physics, Berlin Springer Verlag, ed. {T.~Henning},
  61--141

\bibitem[{{Gail} \& {Sedlmayr}(1988)}]{gail88}
{Gail}, H.-P. \& {Sedlmayr}, E. 1988, \aap, 206, 153

\bibitem[{{Gail} \& {Sedlmayr}(1999)}]{gail99}
{Gail}, H.-P. \& {Sedlmayr}, E. 1999, \aap, 347, 594

\bibitem[{{Gauger} {et~al.}(1990){Gauger}, {Sedlmayr}, \& {Gail}}]{gaug90}
{Gauger}, A., {Sedlmayr}, E., \& {Gail}, H.-P. 1990, \aap, 235, 345

\bibitem[{{Gautschy-Loidl} {et~al.}(2004){Gautschy-Loidl}, {H{\"o}fner},
  {J{\o}rgensen}, \& {Hron}}]{gloidl04}
{Gautschy-Loidl}, R., {H{\"o}fner}, S., {J{\o}rgensen}, U.~G., \& {Hron}, J.
  2004, \aap, 422, 289

\bibitem[{{Grevesse} \& {Anders}(1989)}]{andgrev89}
{Grevesse}, N. \& {Anders}, E. 1989, in American Institute of Physics
  Conference Series, Vol. 183, Cosmic Abundances of Matter, ed.
  {C.~J.~Waddington}, 1--8

\bibitem[{{Grevesse} \& {Sauval}(1994)}]{saugrev94}
{Grevesse}, N. \& {Sauval}, A.~J. 1994, in Lecture Notes in Physics, Berlin
  Springer Verlag, Vol. 428, IAU Colloq. 146: Molecules in the Stellar
  Environment, ed. U.~G. {Jorgensen}, 196

\bibitem[{{Helling} \& {Woitke}(2006)}]{hewo06}
{Helling}, C. \& {Woitke}, P. 2006, \aap, 455, 325

\bibitem[{{Henning} {et~al.}(1995){Henning}, {Begemann}, {Mutschke}, \&
  {Dorschner}}]{hen95}
{Henning}, T., {Begemann}, B., {Mutschke}, H., \& {Dorschner}, J. 1995, \aaps,
  112, 143

\bibitem[{{Herwig}(2005)}]{herw05}
{Herwig}, F. 2005, \araa, 43, 435

\bibitem[{{H{\"o}fner}(2008)}]{hof08bg}
{H{\"o}fner}, S. 2008, \aap, 491, L1

\bibitem[{{H{\"o}fner}(2009)}]{hof09}
{H{\"o}fner}, S. 2009, in Astronomical Society of the Pacific Conference
  Series, Vol. 414, Cosmic Dust - Near and Far, ed. {T.~Henning, E.~Gr{\"u}n,
  \& J.~Steinacker}, 3

\bibitem[{{H{\"o}fner} {et~al.}(2003){H{\"o}fner}, {Gautschy-Loidl}, {Aringer},
  \& {J{\o}rgensen}}]{hof03}
{H{\"o}fner}, S., {Gautschy-Loidl}, R., {Aringer}, B., \& {J{\o}rgensen}, U.~G.
  2003, \aap, 399, 589

\bibitem[{{J{\"a}ger} {et~al.}(2003){J{\"a}ger}, {Dorschner}, {Mutschke},
  {Posch}, \& {Henning}}]{jag03}
{J{\"a}ger}, C., {Dorschner}, J., {Mutschke}, H., {Posch}, T., \& {Henning}, T.
  2003, \aap, 408, 193

\bibitem[{{Jeong} {et~al.}(2003){Jeong}, {Winters}, {Le Bertre}, \&
  {Sedlmayr}}]{jeong03}
{Jeong}, K.~S., {Winters}, J.~M., {Le Bertre}, T., \& {Sedlmayr}, E. 2003,
  \aap, 407, 191

\bibitem[{{Karovicova} {et~al.}(2011){Karovicova}, {Wittkowski}, {Boboltz},
  {Fossat}, {Ohnaka}, \& {Scholz}}]{kar11}
{Karovicova}, I., {Wittkowski}, M., {Boboltz}, D.~A., {et~al.} 2011, \aap, 532,
  A134

\bibitem[{{Koike} {et~al.}(1995){Koike}, {Kaito}, {Yamamoto}, {Shibai},
  {Kimura}, \& {Suto}}]{koi95cor}
{Koike}, C., {Kaito}, C., {Yamamoto}, T., {et~al.} 1995, \icarus, 114, 203

\bibitem[{{Lamers} \& {Cassinelli}(1999)}]{lamcass99}
{Lamers}, H.~J.~G.~L.~M. \& {Cassinelli}, J.~P. 1999, {Introduction to Stellar
  Winds}, ed. {Lamers, H.~J.~G.~L.~M.~\& Cassinelli, J.~P.}

\bibitem[{{Lattimer} \& {Grossman}(1978)}]{latt78}
{Lattimer}, J.~M. \& {Grossman}, L. 1978, Moon and Planets, 19, 169

\bibitem[{{Le Bertre} \& {Winters}(1998)}]{berwin98}
{Le Bertre}, T. \& {Winters}, J.~M. 1998, \aap, 334, 173

\bibitem[{{Mattsson} \& {H{\"o}fner}(2011)}]{matt11}
{Mattsson}, L. \& {H{\"o}fner}, S. 2011, \aap, 533, A42

\bibitem[{{McDonald} {et~al.}(2010){McDonald}, {Sloan}, {Zijlstra},
  {Matsunaga}, {Matsuura}, {Kraemer}, {Bernard-Salas}, \& {Markwick}}]{ian10}
{McDonald}, I., {Sloan}, G.~C., {Zijlstra}, A.~A., {et~al.} 2010, \apjl, 717,
  L92

\bibitem[{{Molster} {et~al.}(2010){Molster}, {Waters}, \& {Kemper}}]{agbgrain}
{Molster}, F.~J., {Waters}, L.~B.~F.~M., \& {Kemper}, F. 2010, in Lecture Notes
  in Physics, Berlin Springer Verlag, ed. {T.~Henning}, Vol. 815, 143--201

\bibitem[{{Norris} {et~al.}(2012){Norris}, {Tuthill}, {Ireland}, {Lacour},
  {Zijlstra}, {Lykou}, {Evans}, {Stewart}, \& {Bedding}}]{norr12}
{Norris}, B.~R.~M., {Tuthill}, P.~G., {Ireland}, M.~J., {et~al.} 2012, \nat,
  484, 220

\bibitem[{{Nowotny} {et~al.}(2011){Nowotny}, {Aringer}, {H{\"o}fner}, \&
  {Lederer}}]{now11}
{Nowotny}, W., {Aringer}, B., {H{\"o}fner}, S., \& {Lederer}, M.~T. 2011, \aap,
  529, A129

\bibitem[{{Nowotny} {et~al.}(2010){Nowotny}, {H{\"o}fner}, \&
  {Aringer}}]{now10}
{Nowotny}, W., {H{\"o}fner}, S., \& {Aringer}, B. 2010, \aap, 514, A35

\bibitem[{{Ordal} {et~al.}(1988){Ordal}, {Bell}, {Alexander}, {Newquist}, \&
  {Querry}}]{ord88fe}
{Ordal}, M.~A., {Bell}, R.~J., {Alexander}, Jr., R.~W., {Newquist}, L.~A., \&
  {Querry}, M.~R. 1988, \ao, 27, 1203

\bibitem[{{Palik}(1985)}]{pal85}
{Palik}, E.~D. 1985, {Handbook of optical constants of solids}, ed. {Palik,
  E.~D.}

\bibitem[{{Pegourie}(1988)}]{peg88sic}
{Pegourie}, B. 1988, \aap, 194, 335

\bibitem[{{Rouleau} \& {Martin}(1991)}]{roul91amc}
{Rouleau}, F. \& {Martin}, P.~G. 1991, \apj, 377, 526

\bibitem[{{Sacuto} {et~al.}(2011){Sacuto}, {Aringer}, {Hron}, {Nowotny},
  {Paladini}, {Verhoelst}, \& {H{\"o}fner}}]{sac11}
{Sacuto}, S., {Aringer}, B., {Hron}, J., {et~al.} 2011, \aap, 525, A42

\bibitem[{{Winters} {et~al.}(2000){Winters}, {Le Bertre}, {Jeong}, {Helling},
  \& {Sedlmayr}}]{wint00}
{Winters}, J.~M., {Le Bertre}, T., {Jeong}, K.~S., {Helling}, C., \&
  {Sedlmayr}, E. 2000, \aap, 361, 641

\bibitem[{{Wittkowski} {et~al.}(2007){Wittkowski}, {Boboltz}, {Ohnaka},
  {Driebe}, \& {Scholz}}]{witt07}
{Wittkowski}, M., {Boboltz}, D.~A., {Ohnaka}, K., {Driebe}, T., \& {Scholz}, M.
  2007, \aap, 470, 191

\bibitem[{{Woitke}(2006)}]{woi06fe}
{Woitke}, P. 2006, \aap, 460, L9

\bibitem[{{Wood}(1979)}]{woo79}
{Wood}, P.~R. 1979, \apj, 227, 220

\bibitem[{{Zeidler} {et~al.}(2011){Zeidler}, {Posch}, {Mutschke}, {Richter}, \&
  {Wehrhan}}]{zeid11}
{Zeidler}, S., {Posch}, T., {Mutschke}, H., {Richter}, H., \& {Wehrhan}, O.
  2011, \aap, 526, A68

\end{thebibliography}

\Online
\begin{appendix} 
\section{Radiative acceleration of the wind}
\label{a_radacc}
The collective opacity for dust particles of radius $\ag$ is given by
\begin{equation}
\kap(\lambda,\ag) = \frac{\pi}{\rho} \ag^2\qq(\lambda,\ag)n_{\mathrm{gr}}=\frac{\pi}{\rho} \frac{\qq(\lambda,\ag)}{\ag}\ag^3n_{\mathrm{gr}},
\end{equation}
where $n_{\mathrm{gr}}$ is the number density of grains in the atmosphere, $\qq$ is the efficiency and $\rho$ is the total mass density. The factor $\ag^3n_{\mathrm{gr}}$ is related to the fraction of atmospheric volume occupied by dust particles. This quantity can be formulated in terms of the volume of a monomer, the basic building block of the grain material, 
\begin{equation}
\ag^3n_{\mathrm{gr}} = \frac{3}{4\pi}V_{\mathrm{mon}}n_{\mathrm{mon}}= \frac{3}{4\pi}\frac{A_{\mathrm{mon}}m_{p}}{\rho_{\mathrm{gr}}}\frac{\varepsilon_{\mathrm{lim}}}{s}n_{\mathrm{H}}f_\mathrm{c}.
\end{equation}
Here we have expressed the volume of the monomer $V_{\mathrm{mon}}$ in terms of the atomic weight of the monomer $A_{\mathrm{mon}}$, the proton mass $m_p$ and the bulk density $\rho_{\mathrm{gr}}$. The number of monomers contained in all grains in a volume of atmosphere $n_{\mathrm{mon}}$ can be expressed by the abundance of the limiting element $\varepsilon_{\mathrm{lim}}$, the number of corresponding atoms in the monomer $s$, the fraction of the limiting element condensed into dust particles $f_{\mathrm{c}}$ and the total number density of H atoms $n_{\mathrm{H}}$. Using $n_{\mathrm{H}}=\rho/(1+4\varepsilon_{\mathrm{He}})$, we obtain Eq. (\ref{e_kap2}) for the dust opacity
\begin{equation}
\kap(\lambda,\ag)=\frac{3}{4}\frac{A_{\mathrm{mon}}}{\rho_{\mathrm{gr}}}\frac{\qq(\lambda,\ag)}{\ag}\frac{\varepsilon_\mathrm{lim}}{s(1+4\varepsilon_{\mathrm{He}})} f_\mathrm{c}.
\end{equation}
\end{appendix}

\begin{appendix} 
\section{Condensation distance}
\label{a_rcond}
The grain temperature $T_{\mathrm{d}}$ is determined by the condition of radiative equilibrium (see Sect. \ref{s_dynmod}),
\begin{equation}
\kappa_{\mathrm{abs,J}}J-\kappa_{\mathrm{abs,S}}S(T_{\mathrm{d}})=0.
\end{equation}
Assuming an optically thin atmosphere, where the incident intensity on the grains is direct star light, the mean intensity $J_{\lambda}$ can be approximated by a geometrically diluted Planck distribution,
\begin{equation}
J_{\lambda} = W(r)B_{\lambda}(T_*).
\end{equation}
In this expression $W(r)$ is the geometric dilution factor
\begin{equation}
W(r)=\frac{1}{2}\left[1-\sqrt{1-(R_*/r)^2}\right],
\end{equation}
which reduces to $W(r)=(R_*/2r)^2$ for $r\gg R_*$. The condition of radiative equilibrium can thus be reformulated as
\begin{equation}
\kappa_{\mathrm{abs,B}}(T_*)W(r)B_{\lambda}(T_*)=\kappa_{\mathrm{abs,B}}(T_{\mathrm{d}})B_{\lambda}(T_{\mathrm{d}}),
\end{equation}
where the source function $S(T_{\mathrm{d}})$ is given by $B_{\lambda}(T_{\mathrm{d}})$. Assuming $r\gg R_*$ and solving for the grain temperature $(T_{\mathrm{d}})$ gives
\begin{equation}
T_{\mathrm{d}}(r) \approx T_*\left(\frac{R_*}{2r}\right)^{1/2}\left(\frac{\kappa_{\mathrm{abs,B}}(T_*)}{\kappa_{\mathrm{abs,B}}(T_{\mathrm{d}})}\right)^{1/4}.
\end{equation}
If the absorption coefficient can be approximated with a power law function, $\kappa_{\mathrm{abs}}\sim\lambda^{-p}$, the factor with the Planck mean opacities can be simplified accordingly,
\begin{equation}
\left(\frac{\kappa_{\mathrm{abs,B}}(T_*)}{\kappa_{\mathrm{abs,B}}(T_{\mathrm{d}})}\right)=\frac{T_*^p}{T_{\mathrm{d}}^p},
\end{equation}
resulting in the following grain temperature distribution
\begin{equation}
T_{\mathrm{d}}(r) \approx T_*\left(\frac{R_*}{2r}\right)^{-\frac{2}{4+p}}.
\end{equation}
Setting $T_{\mathrm{d}}=T_{\mathrm{c}}$ and solving for $R_\mathrm{c}=r(T_\mathrm{c})$ results in the expression for the condensation distance seen in Eq. (\ref{e_rcond}),
\begin{equation}
\frac{R_\mathrm{c}}{R_*} = \frac{1}{2}\left(\frac{T_\mathrm{c}}{T_*}\right)^{-\frac{4+p}{2}}.
\end{equation}
For more details, see \citet{lamcass99}
\end{appendix}

\begin{appendix} 
\section{Levitation distance}
\label{a_levd}
Conservation of energy, with kinetic energy fully converted into potential energy, for a fluid element of mass $m$, with initial velocity $u_0$ at a distance $R_0$ from the star, results in the following expression for the levitation distance $R_{\mathrm{max}}$,
\begin{eqnarray}
\frac{mu_0^2}{2}-\frac{mM_*G}{R_0} &=& 0-\frac{mM_*G}{R_{\mathrm{max}}}\\
\frac{R_*}{R_{\mathrm{max}}} &=& \frac{R_*}{R_0}-\frac{u_0^2}{u_{\mathrm{esc}}^2}\\
\frac{R_{\mathrm{max}}}{R_*} &=& \frac{R_0}{R_*}\left[1- \frac{R_0}{R_*}\left(\frac{u_{\mathrm{0}}}{u_{\mathrm{esc}}}\right)^2\right]^{-1}
\end{eqnarray}
where the escape velocity at the stellar surface is given by  $u_{\mathrm{esc}}=~(2M_*G/R_*)^{1/2}$.

\end{appendix}

\end{document}